\documentclass[pre,preprintnumbers,twocolumn,floatfix,nofootinbib,a4paper,superscriptaddress]{revtex4-2}

\usepackage{color}
\usepackage{calc}
\usepackage{graphicx}
\usepackage{amssymb,amsmath}
\usepackage{tensor}
\usepackage{bm}
\usepackage{float}
\usepackage{microtype}
\usepackage{booktabs}
\usepackage{times}
\usepackage[varg]{txfonts}
\usepackage[colorlinks, pdfborder={0 0 0}]{hyperref}
\usepackage{subfigure}

\hypersetup{citecolor=blue}

\allowdisplaybreaks
\usepackage[utf8]{inputenc}
\usepackage{ulem}
\usepackage{multirow}
\usepackage{array}    
\usepackage{ulem,cancel}
\usepackage{upgreek} 

\usepackage{comment}
\usepackage[nolist]{acronym}
\usepackage{tikz}
\usepackage{textcomp}
\usepackage[export]{adjustbox}
\usepackage{url}

\usepackage{dcolumn}
\usepackage{bm}
\usepackage[]{xcolor}

\newcommand{\p}{\partial}

\newcommand\vect[1]{\mathbf #1}

\newcommand{\dd}{\mathrm{d}}        
\newcommand{\Rey}{\text{Re}}  
\newcommand{\Rm}{R_\text{m}}  
\newcommand{\Pm}{P_\text{m}}  
\newcommand{\Rmcr}{R_\text{m}^{\text{cr}}}  

\newcommand{\TVVV}{T^{V}_{\ VV}}
\newcommand{\TVVM}{T^{V}_{\ VM}}
\newcommand{\TVMM}{T^{V}_{\ MM}}
\newcommand{\TMVM}{T^{M}_{\ VM}}
\newcommand{\TMMM}{T^{M}_{\ MM}}
\newcommand{\bkpq}{b_{kpq}}
\newcommand{\ckpq}{c_{kpq}}
\newcommand{\hkpq}{h_{kpq}}
\newcommand{\thkpq}{\theta_{kpq}(t)}
\newcommand{\mukpq}{\mu_{kpq}}
\newcommand{\muk}{\mu_{k}}
\newcommand{\mup}{\mu_{p}}
\newcommand{\muq}{\mu_{q}}
\newcommand{\const}{\rm const}

\newcommand{\intDk}{\int_{\Delta_k}}

\newcommand{\kmin}{k_{\rm min}}
\newcommand{\kmax}{k_{\rm max}}
\newcommand{\kpeak}{k_{\text peak}}
\newcommand{\bsat}{B_{\rm sat}}
\newcommand{\usat}{u_{\rm sat}}

\newcommand{\Gasym}{\gamma_\text{asym}} 
\newcommand{\Reasym}{\Rey^{\text asym}}
\newcommand{\Rmasym}{\Rm^{\text asym}}

\newcommand{\Fig}[1]{Fig.~\ref{#1}}

\definecolor{darkgreen}{rgb}{0.0, 0.5, 0.0}
\definecolor{doubt}{rgb}{1.0, 0.4, 0.2}

\definecolor{mypurple}{rgb}{0.7,0.3,0.8}

\def\r2#1{\textbf{\textcolor{darkgreen}{#1}}} 

\usetikzlibrary{arrows,shapes,trees,decorations.pathreplacing}
\usepackage{aas_macros}
\usepackage{diagbox}

\allowdisplaybreaks

\begin{document}


\title{Small-scale dynamo saturation across magnetic Prandtl numbers using the EDQNM closure}

\author{Muhammed Irshad}
\email{muhammed.irshad@icts.res.in}
\affiliation{International Centre for Theoretical Sciences, Tata Institute of Fundamental Research, Bangalore 560089, India}
\author{Kandaswamy Subramanian}
\affiliation{Inter-University Centre for Astronomy and Astrophysics, Post Bag 4, Ganeshkhind, Pune 411007, India}
\affiliation{Department of Physics, Ashoka University, Rajiv Gandhi Education City, Rai, Sonipat 131029, India}
\author{Pallavi Bhat}
\affiliation{International Centre for Theoretical Sciences, Tata Institute of Fundamental Research, Bangalore 560089, India}

\begin{abstract}
Small-scale dynamos {(SSDs)} are believed to be the primary source of magnetic fields in 
{all turbulent astrophysical systems, especially those with weak rotation} 
such as elliptical galaxies and galaxy clusters.  
{The initial kinematic {phase} of these dynamos} is relatively well understood.
{Here} we demonstrate analytically and numerically that, in an appropriate limit, the {eddy-damped quasi-normal Markovian (EDQNM) closure for incompressible magnetohydrodynamic turbulence is strictly equivalent to the earlier models {of} kinematic dynamos.} 
{Moreover, it allows the extension of the {kinematic} dynamo} framework {to multi-scale turbulent flows and} 
into the nonlinear regime. 
The EDQNM closure also enables us to explore a wide parameter range which is inaccessible to direct numerical simulations {of the SSD}.
Using nonhelical EDQNM simulations, we identify several asymptotic regimes of nonlinear dynamo action {when the system is highly turbulent with {fluid Reynolds number} $\Rey \gtrsim 10^6$ for {magnetic Prandtl number} $\Pm > 1$ and {magnetic Reynolds number} $\Rm \gtrsim 10^6$ for $\Pm < 1$}:
1) 
the kinematic growth rate approaches a value independent of $\Pm$, 
2) the saturated magnetic to kinetic energy ratio similarly converges to $\simeq 0.55$ {across $\Pm$}, while the ratio of magnetic to kinetic integral wavenumbers asymptotes to $\simeq 3$. For {all} {$\Pm$}, we further find strong feedback {between magnetic field and velocity field {largely} via Alfv\'{e}nisation leading to a saturated kinetic and magnetic spectra with almost the same inertial range with a slope of $-3/2$.} 
{These findings could provide guidance for future {global} simulations and for modeling the nonlinear regime of astrophysical systems living in these extreme limits.}
\end{abstract}

\maketitle



\section{Introduction}
Observations reveal the {ubiquitous} presence of magnetic fields 
across astrophysical systems, from planets, stars to galaxies and galaxy clusters. They are thought to 
arise from turbulent dynamo action, whereby the random motions in a highly conducting plasma
amplifies a seed magnetic field and then maintains it against decay. In many systems both random fields
on scales smaller than the turbulent energy injection scale, and more coherent larger scale 
fields are detected
\citep{Clusterfield_2005,reiners_observations_2012, M51_B, beck2015, Vacca2018, hull_interferometric_2019, Beck2020, Borlaff+2023}. 
These small scale fields are believed to arise from the small-scale dynamo (SSD), which operates in any sufficiently conducting turbulent plasma 
whereas the amplification of a large-scale field requires 
the presence
of either kinetic or magnetic helicity in the turbulence \citep{AxelKandu, Rincon_2019, Tobias_2021, shukurovkandubook}.
In systems with negligible rotation, such as galaxy clusters and elliptical galaxies, the observed magnetization is therefore thought to arise primarily from the SSD \citep{Moss_Shukurov_1996, KS2006Cluster, PB_Faradayrot, Sur_2019, Seta_2021_ellgal}. 
Moreover, this can generate strong magnetic fields already during structure formation \cite{irshad}. 
Beyond its astrophysical relevance, SSD also provides an opportunity to understand the forced magnetohydrodynamic (MHD) turbulence \cite{Schekochihin_2022}.

The kinematic stage of the small-scale dynamo, where Lorentz force is negligible, is well understood theoretically \citep{kazantsev, KN1967,KA92, subramanian1997dynamics}, in more realistic regimes \citep{BS14,Bhat_Subramanian_2015} and numerically \citep{Haugen2004, Schekochihin_2004}. 
In any sufficiently conducting plasma flows with magnetic Reynolds number ($\Rm$) greater than a critical value ($\Rmcr$), magnetic energy grows exponentially and the magnetic energy spectrum peaks near the resistive scale. As the magnetic field amplifies, the Lorentz force back reacts on the flow, eventually saturating the magnetic energy. 
However, this nonlinear stage remains less understood, with debates on the saturated magnetic field morphology, spectra and how the Lorentz force reorders the field.

A theoretical model \citep{Kandu_unified, Subramanian2003Hyperdiffusion} incorporates magnetic back reaction through an ambipolar{-type} drift velocity {due to the } Lorentz force and suggest{s} that nonlinear effects enhance effective resistive dissipation, thereby reducing $\Rm$ toward $\Rmcr$ and leading to saturation. In this picture, the saturated magnetic energy spectrum resembles the critical eigenfunction of the kinematic dynamo, {with the power} shifted {from resistive scales} to larger scales, 
a fraction $1/\sqrt{\Rmcr}$ of the {forcing scale}.
{A different model applicable to highly viscous flows 
\cite{KIM} argues that Lorentz forces lead to suppression of
random stretching by the flow leading to saturation.} 
Whereas \cite{Schekochihin_2002} propose that magnetic fields remain concentrated at the resistive scales and reorganize toward the viscous scale only on resistive timescales, thereby retaining a predominantly small-scale spectrum. These and other attempts \citep{SetaBhatSubramanian2015, Xu_2016} to theoretically understand the SSD nonlinear regime have remained inconclusive.

Direct numerical simulations (DNS) {in 3-D} have greatly advanced our understanding of the nonlinear SSD \citep{Meneguzzi, Haugen2004, Schekochihin_2004, Schekochihin_2007, Sahoo_2011, Brandenburg_2011, Seta2020_sat, 10.1093/mnras/stad3535, Warnecke_2023, Beattie2023, Kriel2026, Gent_2026, Beresnyak2012}. 
They demonstrate a universal linear growth of magnetic energy before saturation {in subsonic flows} and the ordering of magnetic fields to large scales. Simulations of \cite{Galishnikova2022} suggests the importance of 
{tearing-mediated reconnection} in turbulent dynamo saturation.  Yet, there is no consensus on the characteristic scale of the saturated magnetic energy spectrum nor on the {mechanism of} how the Lorentz force saturates the magnetic field {growth}. Furthermore, most DNS have focused on {magnetic Prandtl number $\Pm = \Rm / \Rey$} $\gtrsim 1$ relevant to interstellar and intracluster media, while regimes with $\Pm \ll 1$ common in planetary and stellar interiors remain comparatively less explored.
A major bottleneck is the high computational cost of DNS, which limits access to the extreme kinetic and magnetic Reynolds numbers ($\Rey$, $\Rm$) and 
{magnetic Prandtl numbers  ($\Pm$)} typical of astrophysical systems.

In this work, we employ the eddy-damped quasi-normal Markovian (EDQNM) closure, which has been successfully used to study MHD turbulence and dynamo processes \citep{PFL1976, Leorat_1981, Grappin1982, Murugan2026}. 
We are in addition motivated by several important considerations. First, as we
show {below}, the EDQNM formalism, in an appropriate limit, recovers the classical kinematic dynamo results of \citep{kazantsev, KN1967, KA92}, thereby establishing their equivalence. 
Moreover, in this limit, when magnetic back reaction is partially incorporated, 
{it} leads to results comparable to the earlier analytic models. {Then} it
appears 
that the EDQNM subsumes many previous approaches. 
{Furthermore it} 
allows us to explore regimes of large $\Rey$, $\Rm$, including both $\Pm \gg 1$ and $\Pm \ll 1$ at a significantly reduced computational cost compared to DNS. 

The paper is organized as follows. Section~\ref{sec:review} reviews the current understanding of kinematic and nonlinear SSD. Section~\ref{sec:closure} presents the EDQNM equations and section~\ref{sec:edqnm_limit} analytically and numerically shows their equivalence to previous theories of  kinematic dynamo in a particular limit. 
{It also} discuss{es} the impact of nonlinear mode coupling on the saturation of magnetic energy spectrum. Section~\ref{sec:edqnm_sim} describes the results from numerical simulations of EDQNM, including the asymptotic regimes identified for kinematic growth rate, saturation efficiency, ratio of magnetic to kinetic 
{integral wavenumbers}
and the evolution of respective
{energy spectra}. Finally section~\ref{sec:discussion} 
{presents a discussion of} the important results and 
their  implications 
{along with our conclusions}.

\section{Small scale dynamos}\label{sec:review}
The evolution of an incompressible conducting plasma is governed by the equations of \textit{magnetohydrodynamics} (MHD)
\begin{gather}
\frac{\text{D} \vect{v} }{\text{D} t} 
   = -\frac{\nabla p}{\rho} + \vect{F} +  \frac{\left(\nabla\times\vect{B}\right)\times\vect{B}}{4\pi\rho}
+ \nu \nabla^2 \vect{v},\quad \nabla\cdot\vect{v} = 0,\label{eq:mhd1b}\\
\frac{\p \vect{B}}{\p t} = \nabla\times\left(\vect{v}\times \vect{B}\right) + \eta \nabla^{2}\vect{B}, \quad \nabla\cdot\vect{B} = 0, \label{eq:mhd1c}
\end{gather}
where $\text{D}/\text{D}t= \partial/\partial t + (\vect{v}\cdot\nabla),$ plasma velocity is $\vect{v}$, with pressure $p$, density $\rho$, 
$\vect{B}$ is the magnetic field, $\nu$ and $\eta$ are the kinematic viscosity and magnetic diffusivity, respectively. The turbulence in the plasma is driven by forcing $\vect{F}$ around the wavenumber $k_f$.
The system is characterized by the kinetic and magnetic Reynolds numbers $\Rey = V/k_f\nu,\ \Rm = V / k_f\eta$. Here $V$ is the characteristic speed.
The magnetic Prandtl number $\Pm = \Rm / \Rey$ determines the separation between viscous ($k_\nu$) and resistive ($k_\eta$) {wavenumbers}. 
{For $\Pm > 1$, these are related as $k_\eta = k_\nu \Pm^{1/2}$ and at $\Pm < 1$, $k_\eta = k_\nu \Pm^{3/4}$ assuming Kolmogorov kinetic energy spectrum.} 

In the kinematic stage of the SSD, magnetic field is sufficiently weak that the Lorentz force is negligible. 
This decouples Eq.~\eqref{eq:mhd1c} from Eq.~\eqref{eq:mhd1b}, allowing the magnetic field to be solved for a prescribed flow. The first successful theoretical treatment of this kind was done by \cite{kazantsev} by solving for the magnetic correlation function using $\updelta$-correlated {in time}, Gaussian random velocity field. Around the same time, \cite{KN1967} derived the evolution equation for the magnetic energy spectra using a different closure and found weak exponential growth. {Equivalent formulations and results were presented later in detail by 
\citep{Vainshtein1983, KA92, subramanian1997dynamics, KimHughes1997}.}

{The formulation by Kazantsev \citep{kazantsev} presents an evolution equation for the magnetic energy spectrum $M(k,t)$ (note that the original equation was derived for the 3D spectrum; see Appendix \ref{subsec:kaz} for details on the conversion to 1D spectrum)} 
\begin{align}\label{eq:kin2}
    \frac{\p M}{\p t} = \int K_m (k,p)\ M(p,t)\ \dd p  - 2\left({\eta + \eta_T}\right) k^2 M(k,t),
\end{align}
where 
\begin{align}\label{eq:kin3}
  K_m (k,p) &= \tau k^4 \int_0^\pi \dd\theta \ \sin^3\theta\ \left(k^2 + p^2 - kp\cos\theta\right)\ \frac{E(q)}{q^4},
\end{align}
with $q^2 = k^2 + p^2 - 2kp \cos\theta$. {Where $\tau$ is the correlation time of the flow. In the case of $\updelta$-correlated flows, $\tau\to 0$ and $E(q) \sim 1/\tau$ to keep $\tau E(q)$ finite.}  
The first term represents the random stretching of magnetic field by the velocity field leading to amplification in magnetic energy,
and the second term dissipates magnetic energy due to microscopic diffusivity $\eta$, and turbulent diffusivity
\begin{align}\label{eq:kin4}
    {\eta_T} &= \frac{2}{3} {\tau} \int E(q)\ \dd q = \frac{1}{3}{\tau} \langle v^2\rangle.
\end{align}
The dynamo is active, when
stretching dominates over the total diffusion and magnetic Reynolds number exceeds a critical value, $\Rm > \Rmcr$.

In the viscous regime ($k \gg k_\nu$) {of high $\Pm$ flows ($k_\nu < k_\eta$)}, Eq.~\eqref{eq:kin2} simplifies to \citep{kazantsev, KA92}
\begin{align}\label{eq:kin5}
    \frac{\p M}{\p t} = \frac{\gamma}{5} \left(k^2 \frac{\p^2 M}{\p k^2} - 2k\frac{\p M}{\p k} + 6M\right) - 2 {\eta k^2}M{,}
\end{align}
{where $\gamma$ is of the order of flow shear rate given by
\begin{equation}\label{eq:gamma}
    \gamma = \frac{1}{3}{\tau}\int k^2 E(k)\, \dd k,
\end{equation}
and proportional to the growth rate of magnetic energy \citep{subramanian1997dynamics, Kandu_unified, Schober_2012, Bovino_2013}.}
In the initial stage of the SSD, weak seed field on large scales   $k' \ll k_\eta$ gets stretched by the velocity field and the peak of ME spectrum moves to resistive scale.
During this period magnetic spectrum leaves behind $M(k) \propto k^{3/2}$ between $k_\nu$ and $k_\eta$. Then onwards, it follows the self similar evolution
\begin{align}\label{eq:kin7}
    M(k, t) \propto \ k^{3/2}K_0(k/k_\eta)\ \exp{\left(3\gamma t / 4\right)},
\end{align}
where $K_0$ is the Macdonald function. 

By the end of the kinematic stage, dynamically strong magnetic field gets developed and the Lorentz force couples Eqs~\eqref{eq:mhd1b} and \eqref{eq:mhd1c}. Thus in general, 3-D direct numerical simulations (DNS) are employed to study the nonlinear dynamos. DNS reveal that in the nonlinear stage, the fastest but weaker eddies cannot stretch the field lines further and the slowest but stronger eddies take over. This may continue until the magnetic energy achieves scale by scale 
{saturation, with the total magnetic energy (ME) a fraction of the kinetic energy (KE).}

\section{Closure model}\label{sec:closure}
The theory described above well explains the kinematic stage of SSD and to understand the nonlinear saturation of magnetic energy, direct numerical simulations are employed. But due to the computational constraints, mostly $\Pm = 1$ simulations are performed in the literature. 
Moreover, exploring $\Pm \ll 1$ and ${\Pm} \gg 1$, with large $\Rey$ and $\Rm$ is almost impossible with the DNS. 
In this section, we introduce the eddy-damped quasi-normal Markovian (EDQNM) closure formulation for incompressible MHD turbulence, which we use in the latter sections to study the SSD across {the parameter space.} 
Also we show that, in an appropriate limit, it recovers the standard kinematic dynamo theory discussed earlier.

In any statistical theory of turbulence, the evolution of lower order moments depends on higher order moments, leading to the well known closure problem. The EDQNM closure involves \cite{Lesieur2008}
\begin{enumerate}
    \item \textit{Quasi - normal:} 
    {In the evolution equation for the third order correlators, the} fourth order correlations are expressed as sum of product of second order moments, 
    {and an irreducible part.} 
    \item \textit{Eddy damping:} 
    {This irreducible part is replaced by a damping term proportional to the 
    third order correlation, the `eddy damping' term. The resulting solution is substituted back into the evolution for the second order moments.}
    
    \item \textit{Markovianization:} 
    To ensure the positivity of energy spectra, 
    {the eddy damping is assumed so strong that the third order moment responds to the instantaneous
    value of the product of second order moments. The evolution equation for the second order moments then becomes local in time.}
\end{enumerate}
The resulting equations for the kinetic energy spectrum $E_k \equiv E(k,t)$ and magnetic energy spectrum $M_k \equiv M(k,t)$ in the nonhelical MHD turbulence are \citep{PFL1976, Grappin1982}
\begin{align}
    \left(\frac{\p}{\p t} + 2\nu k^2\right) E_k &= F_k + \int_{\Delta_k} \dd p \; \dd q \; \theta_{kpq}\; \left(\TVVV + \TVVM + \TVMM \right), 
    \label{eq:cl1}\\[5pt]
    \left(\frac{\p}{\p t} + 2\eta k^2\right) M_k &= \int_{\Delta_k} \dd p \; \dd q \; \theta_{kpq}\; \left(\TMVM + \TMMM \right).\label{eq:cl1a}
\end{align}
Kinetic energy $E_V$ and magnetic energy $E_M$ are given by the area under their respective spectra.
The flow is driven by a forcing of amplitude $F_0$ around wavenumber $k_f$
\begin{equation}\label{eq:force}
    F_k = F_0\, \frac{k^4 \exp(-2k^2/k_f^2)}{\int k^4 \exp(-2k^2/k_f^2)\, \dd k }.
\end{equation}
And the dissipation is governed by kinematic viscosity $\nu$ and magnetic diffusivity $\eta$.  
The transfer terms represent interactions between velocity and magnetic fields and can be grouped according to the participating fields
\begin{align}\label{eq:cl2}
    \TVVV &= \frac{k}{pq}\ b_{kpq}\ \left(k^2 E_p E_q - p^2 E_q E_k\right), \\[5pt]
    \TVVM &= -\frac{kp}{q}\ c_{kpq}\ E_k M_q, \\[5pt]
    \TVMM &= \frac{k^3}{pq}\ c_{kpq}\ M_p M_q, \\[5pt]
    \TMVM &= \frac{k^5}{p^3q}\ c_{kpq}\ E_p M_q + \frac{k}{pq}\ h_{kpq}\ \left(k^2 M_p E_q - p^2 E_q M_k\right) , \\[5pt]
    \TMMM &= -\frac{k^3}{pq}\ c_{kpq}\ M_q M_k.
\end{align}
The terms $\TVVV$, $\TVVM$,\footnote{ \cite{PFL1976} misses a negative sign in $\TVVM$ as pointed out in the appendix of \cite{Grappin1982}. The correct sign is necessary to ensure total energy conservation. We thank Annick Pouquet for alerting us about this.} and $\TVMM$ contribute to the evolution of kinetic energy, while $\TMVM$ and $\TMMM$ govern the evolution of magnetic energy.
These interactions involve wave vectors satisfying $\vect{k} + \vect{p} + \vect{q} = \vect{0}$, and the integrals are taken over such triangles denoted by $\Delta_k$.
The geometric coefficients in the transfer terms are 
\begin{align}\label{eq:cl3}
    \bkpq &= \frac{p}{k}\left(xy + z^3\right),
    \:
    \ckpq = \frac{pz}{k}(1-y^2),
    \:
    \hkpq = 1 - y^2,
\end{align}
with the direction cosines 
\begin{align}\label{eq:cl4}
    x = \cos{\alpha}, \: y = \cos{\beta}, \: z = \cos{\gamma}, 
\end{align}
where $\alpha,\ \beta$ and $\gamma$ are the angles opposite to the sides $\bm{k},\ 
\bm{p}$ and $\bm{q}$ of the triangle. 

Finally, the triad time relaxation operator obtained from the eddy damping procedure is
\begin{align}\label{eq:cl5}
    \thkpq = \frac{1}{\mukpq} \left(1 - e^{-\mukpq t}\right),
\end{align}
with the damping rate $\mukpq = \muk + \mup + \muq$. In presence of magnetic field the damping rate for each mode is given by 
\begin{align}\label{eq:cl6}
    \muk = \left(\nu + \eta\right) k^2 + C_s \left(\int_0^k \ell^2 \left(E_\ell + M_\ell\right) \ \dd \ell\right)^{1/2} + 
    \notag\\
    \frac{k}{\sqrt{3}} \left(2 \int_0^k M_\ell \; \dd\ell\right)^{1/2},
\end{align}
which includes contributions from viscous and resistive dissipation, nonlinear scrambling of the fields, and Alfv\'{e}nic interactions respectively.
Note that we use $C_s = 0.26$ while performing the simulations as suggested in \cite{Grappin1982}.

\section{EDQNM: $\thkpq = \tau$ limit}\label{sec:edqnm_limit}
We now show that the EDQNM formulation recovers Eq.~\eqref{eq:kin2} of kinematic dynamo, when the magnetic field back reaction is weak and triad relaxation time is a constant, $\thkpq = \tau$. 
\subsection{Kinematic dynamo}
In this limit, since $\TMMM$ is negligible, RHS of Eq.~\eqref{eq:cl1a} becomes
\begin{align}\label{eq:cl7a}
     I  &= \tau \int_{\Delta_k} \dd p\ \dd q\ \Bigg\{\frac{k^5}{p^3q}\ c_{kpq}\ E_p M_q + \frac{k^3}{pq}\ \hkpq\ M_p E_q \notag\\
    & \quad\qquad\qquad\qquad\qquad\qquad\qquad- \frac{kp}{q}\ \hkpq\ E_q M_k\Bigg\} \notag\\
    &= {I_1 + I_2 + I_3}
\end{align}
The last term evaluates to the turbulent diffusion (see {Appendix~\ref{eval-tfr}})
\begin{align}\label{eq:cl7b}
    I_3 = -\tau k M_k\int_{\Delta_k} \dd p\ \dd q\ \frac{p}{q}\ \hkpq\ E_q =   -2\eta_T k^2 M_k,
\end{align}
with the coefficient
\begin{align}\label{eq:cl7c}
    \eta_T = \frac{2}{3} \tau \int E_q\ \dd q = \frac{1}{3} \tau \left<v^2\right>.
\end{align}
And the first two terms combine {to give}, $I_1 + I_2 =\int K_m(k, p)M(p)\ \dd p,$ where 
\begin{align}\label{eq:cl7d}
    K_m (k,p) = \tau k^4\int \dd\theta \ \sin^3\theta\ \left(\frac{k^2 + p^2 - kp\cos\theta}{q^4}\right) \ E_q,
\end{align}
with $q(p,\theta) = \left(k^2 + p^2 - 2kp\cos\theta\right)^{1/2}$ {(see appendix~\ref{eval-tfr})}. Finally we have
\begin{align}\label{eq:cl7}
    \left(\frac{\p}{\p t} + 2\left(\eta + \eta_T\right) k^2\right) M_k &= \int K_m(k, p)\ M(p,t)\ \dd p.
\end{align}
Remarkably, this is identical to Eq.~\eqref{eq:kin2}, 
demonstrating that the EDQNM closure recovers the kinematic dynamo equation 
{derived first by Kazantsev \cite{kazantsev} and also by 
\cite{KA92} for the
$\updelta$-correlated in time flow}. 
We refer to this limit with $\thkpq = \tau$ as the `Kazantsev dynamo' with in the EDQNM framework.

We numerically solve Eq.~\eqref{eq:cl1a} for the ME spectrum with $\tau=1$ in a wavenumber grid spanning from $k_\text{min} = 1$ to $k_{\text max} = 256$. 
Details of our numerical implementation are provided in {Appendix~\ref{num-setup}}. The prescribed kinetic energy spectrum and the initial magnetic energy spectrum share a functional form similar to the forcing function in Eq.~\eqref{eq:force}, peaking at $k=2$, but with respective amplitudes $E_0$ and $M_0$. To isolate the kinematic stage, we neglect the magnetic back-reaction term, $\TMMM$. The relevant parameters are summarized in Table~\ref{tab:kaz_params}.

\begin{table}[htbp]
\centering
\caption{Parameters of the EDQNM simulations of Kazantsev dynamo with $\thkpq = 1$. The forcing wavenumber is fixed at $k_f = 2$  for all runs and we use $k_E = k_p = k_f$.}
\label{tab:kaz_params}
\begin{tabular}{l c c c c c c c c c}
\toprule
Run & $\nu$ & $\eta$ & $F_0$ & $E_0$ & $M_0$ & $u_{\text{sat}}$ & $\Rm$ & $\Pm$ \\
\midrule
Kinematic & -- & $4\times10^{-3}$ & 5.0 & $1.0$ & $10^{-9}$ & 1.43 & 179 & 249 \\
NL & 1.0 & $4\times10^{-3}$ & 10.0 & $1.0$ & $10^{-9}$ & 1.52 & 190  & 250 \\
\bottomrule
\end{tabular}
\end{table}

\begin{figure}
    \centering
    \includegraphics[width=1.0\linewidth]{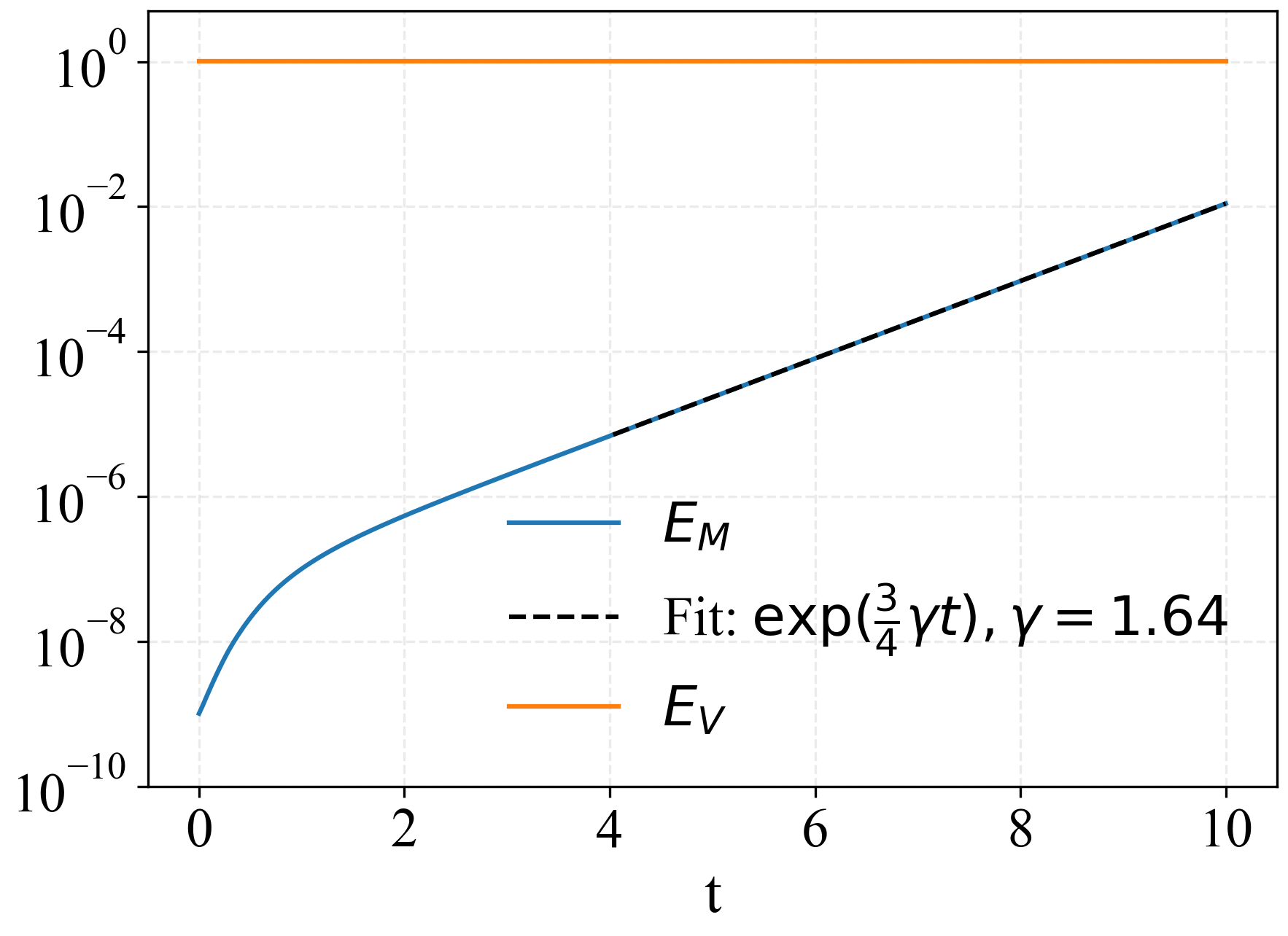}
    \caption{Evolution of kinetic energy (orange line) and magnetic energy (blue line) along with the exponential fit (dashed) in the  kinematic Kazantsev dynamo simulation with $\Rm \simeq 179$. 
    {The $\gamma$ value thus obtained matches fairly well with the theoretical expectation of} {$\gamma = (1/3) \int k^2 E_k \dd k = 1.73$.} 
    }
    \label{fig:Emag_lin512}
\end{figure}

\begin{figure}
    \centering
    \includegraphics[width=\linewidth]{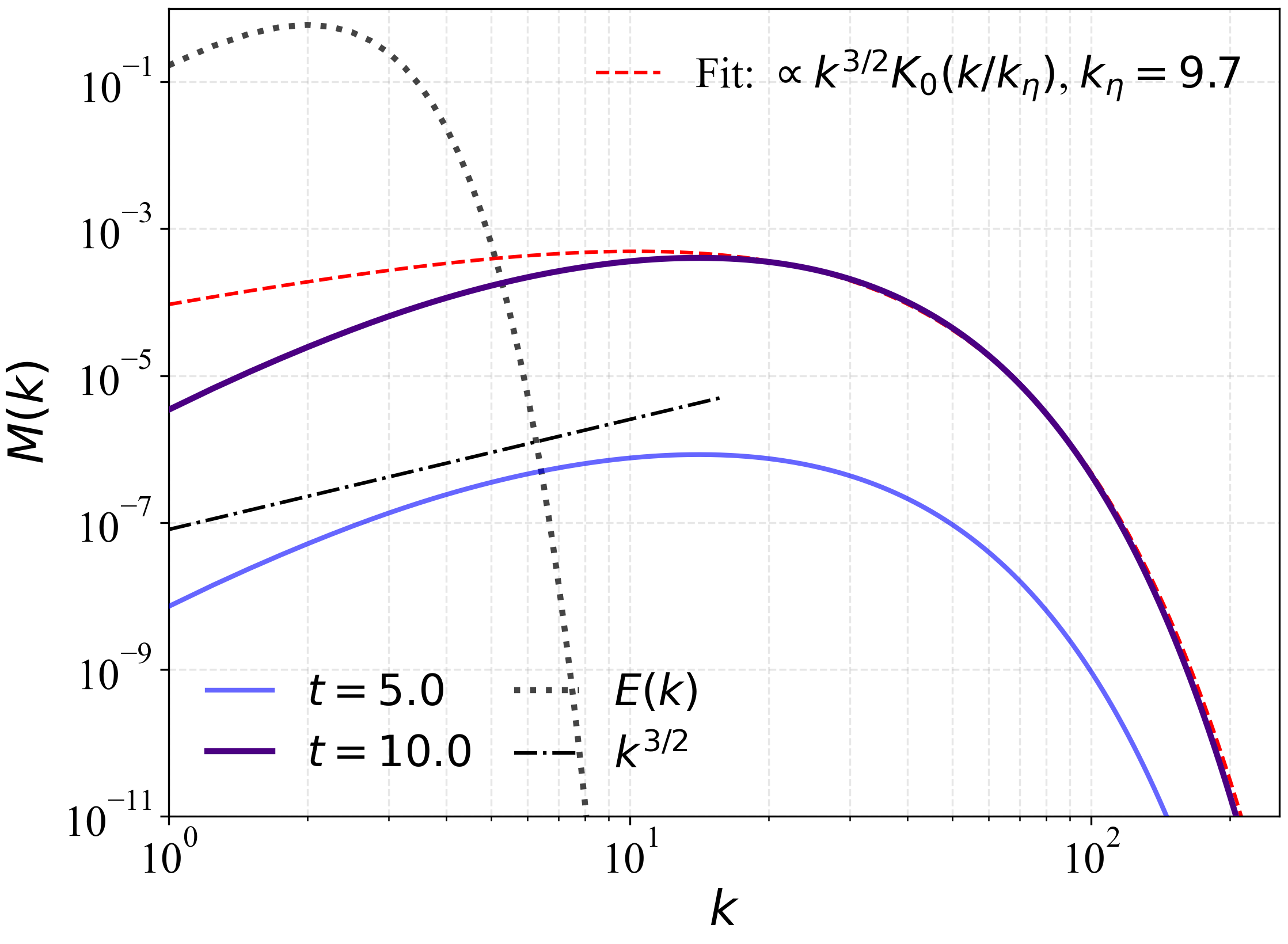}
    \centering
    \caption{Evolution of magnetic energy spectra in the kinematic Kazantsev dynamo is shown along with the kinetic energy spectrum (dotted). Dash-dotted line is the Kazantsev spectra $k^{3/2}$ for reference. Dashed line shows the fit using Eq.~\eqref{eq:kin7}, which fits the smaller scales well.
    }
    \label{fig:spec_lin256_kf2}
\end{figure}

Figure~\ref{fig:Emag_lin512} shows the exponential growth of the magnetic energy (blue) following an initial transient, along with a numerical fit (dashed). 
The analytical prediction for the growth rate, $\gamma = (1/3) \int k^2 E_k \dd k = 1.73$, agrees fairly well with the fitted value of 1.64. This slight discrepancy may be attributed to numerical errors accumulated by the solver or inaccuracies in evaluating the transfer terms. The kinetic energy remains constant, as we do not solve Eq.~\eqref{eq:cl1} during this kinematic stage.

The evolution of the ME spectrum in Fig.~\ref{fig:spec_lin256_kf2} initially shows {significant amplification on smaller scales,} 
accompanied by the migration of its peak ($k_p$) from larger scales toward the resistive scales. 
Subsequently, the spectrum grows self-similarly, showing an excellent fit to the Kazantsev eigenfunction, Eq.~\eqref{eq:kin7}, at small scales. Deviations at large scales arise from 
the assumptions used to derive Eq.~\eqref{eq:kin5}. 
These results {numerically} demonstrate that the EDQNM closure with a constant relaxation time quantitatively reproduces the predictions of kinematic dynamo theory 
{of \citep{kazantsev,KA92}.}

\subsection{Nonlinear Kazantsev dynamo}
\begin{figure}
    \centering
    \includegraphics[width=\linewidth]{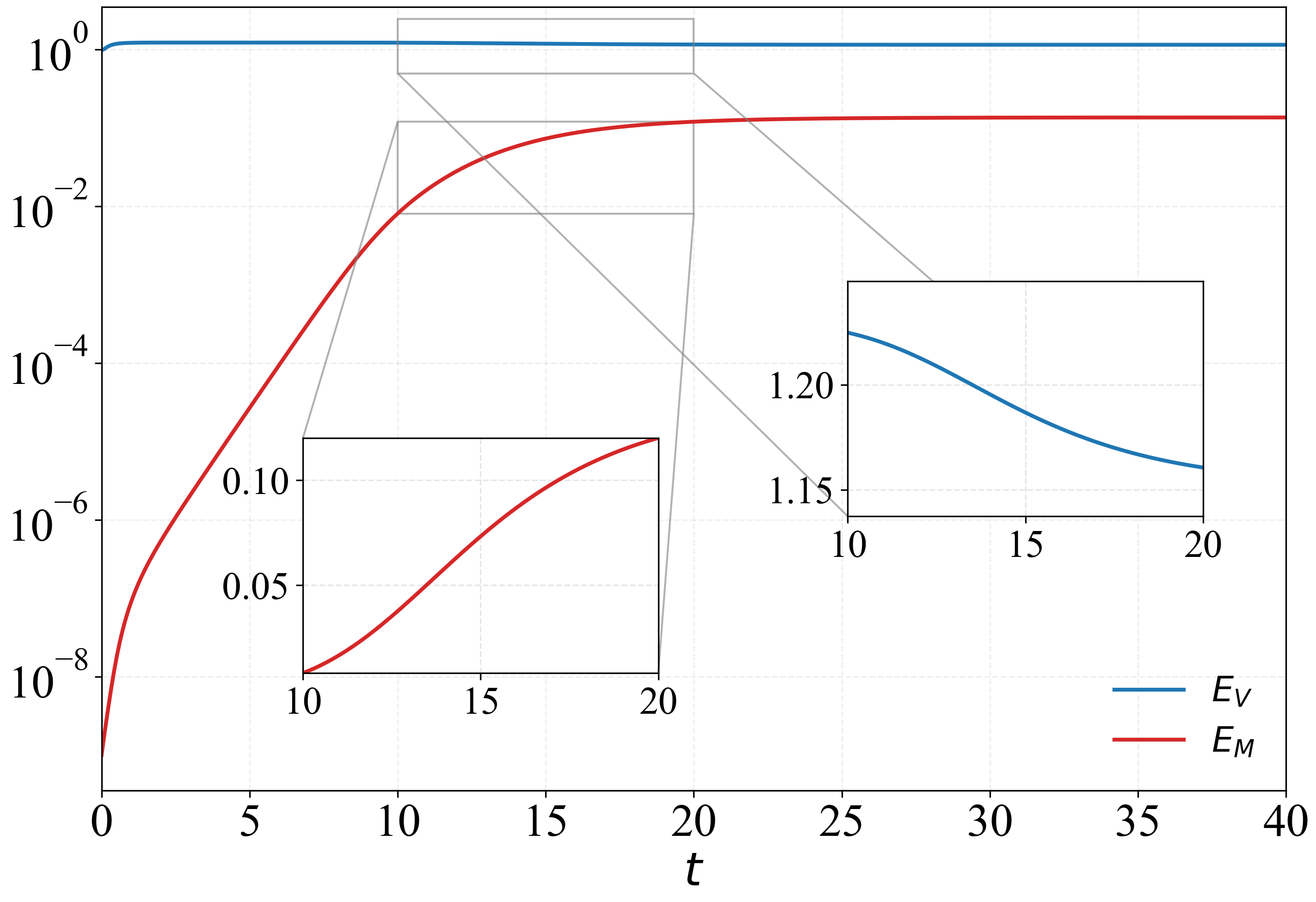}
    \caption{Evolution of ME (red curve) and KE (blue curve) in nonlinear Kazantsev dynamo. The insets focus on the linear growth in ME and reduction in KE during the nonlinear stage. 
    }
    \label{fig:Emag-kazfull}
\end{figure}

We now solve the full EDQNM Eqs~\eqref{eq:cl1}~–~\eqref{eq:cl1a} but
with $\thkpq = \tau = 1$ to investigate the nonlinear saturation of ME. 
{We continue to call this the `Kazantsev dynamo' albeit the nonlinear version as we are still in the limiting case of $\thkpq = \const$.}
As shown in Fig.~\ref{fig:Emag-kazfull}, magnetic energy grows 
exponentially in the kinematic stage. Once the Lorentz force back reaction on the flow is significant, nonlinear stage ensues and the ME grows linearly (see inset {of \Fig{fig:Emag-kazfull}}). Finally, the ME saturates in sub-equipartition with KE. 
The nonlinear stage also shows a reduction in kinetic energy as shown in the inset.

\begin{figure}
    \centering
    \includegraphics[width=\linewidth]{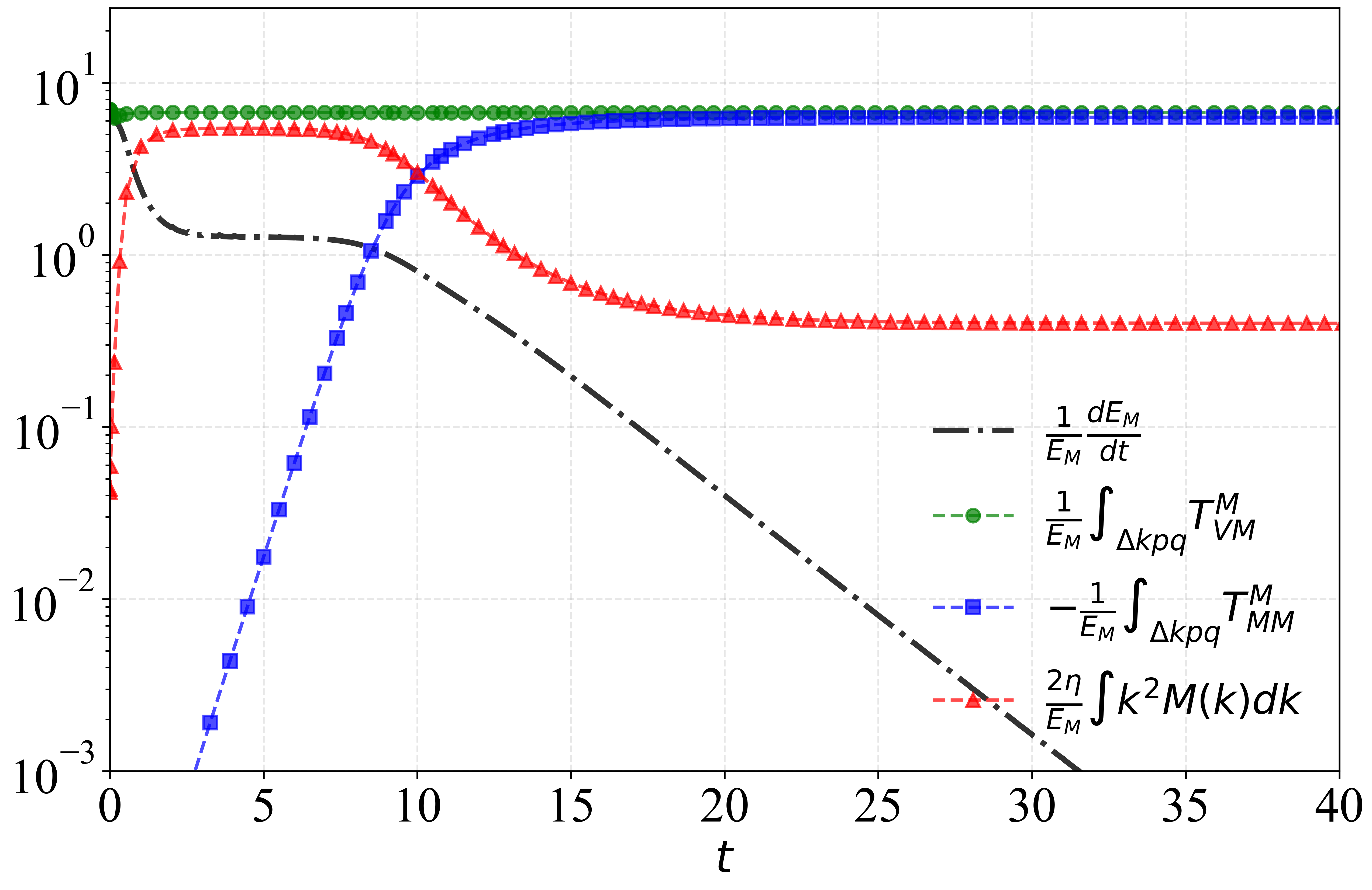}
    \caption{Term by term comparison of Eq.~\eqref{eq:cl1a} integrated over $k$ and normalized by the instantaneous magnetic energy in the nonlinear Kazantsev dynamo. Note that $\int_{\Delta kpq} \TMMM$ is always negative and $\int_{\Delta kpq} \TMVM$ is a combination of stretching and turbulent dissipation as derived in Eqs~\eqref{eq:cl7a} - \eqref{eq:cl7}.}
    \label{fig:magterms_vs_t_fullkaz}
\end{figure}

The mechanism responsible for {the} saturation {of magnetic energy} is revealed by a term-by-term 
analysis of Eq.~\eqref{eq:cl1a} {integrated over $k$}. 
{The terms are normalized by the magnetic energy $E_M(t)$ so their values represent rates (i.e., inverse timescales) associated with the corresponding processes} {and shown in Fig.~\ref{fig:magterms_vs_t_fullkaz}.} 
{For simplicity, we use the notation: $\int_{\Delta kpq} \equiv \int \dd k\, \intDk \thkpq\, \dd p\, \dd q$.}
In the kinematic stage, the dynamo is active owing to a slight excess of stretching over microscopic and turbulent dissipation 
{(the green dots at a slightly higher amplitude compared to the red triangles)}. Note that $\int_{\Delta kpq} \TMVM$ is a combination of stretching and turbulent dissipation as derived in Eqs~\eqref{eq:cl7a} - \eqref{eq:cl7}.
In the nonlinear stage, 
{the interaction among different modes of the magnetic field yields a significant contribution in} $\int_{\Delta kpq} \TMMM$ (blue squares){. This} term is always negative {and its magnitude exceeds the microscopic resistive dissipation by nearly two orders of magnitude} 
{providing an} effective {sink of magnetic energy} to balance the stretching and halt further growth {in} ME, {leading to saturation}. 

\begin{figure}
    \centering
    \includegraphics[width=\linewidth]{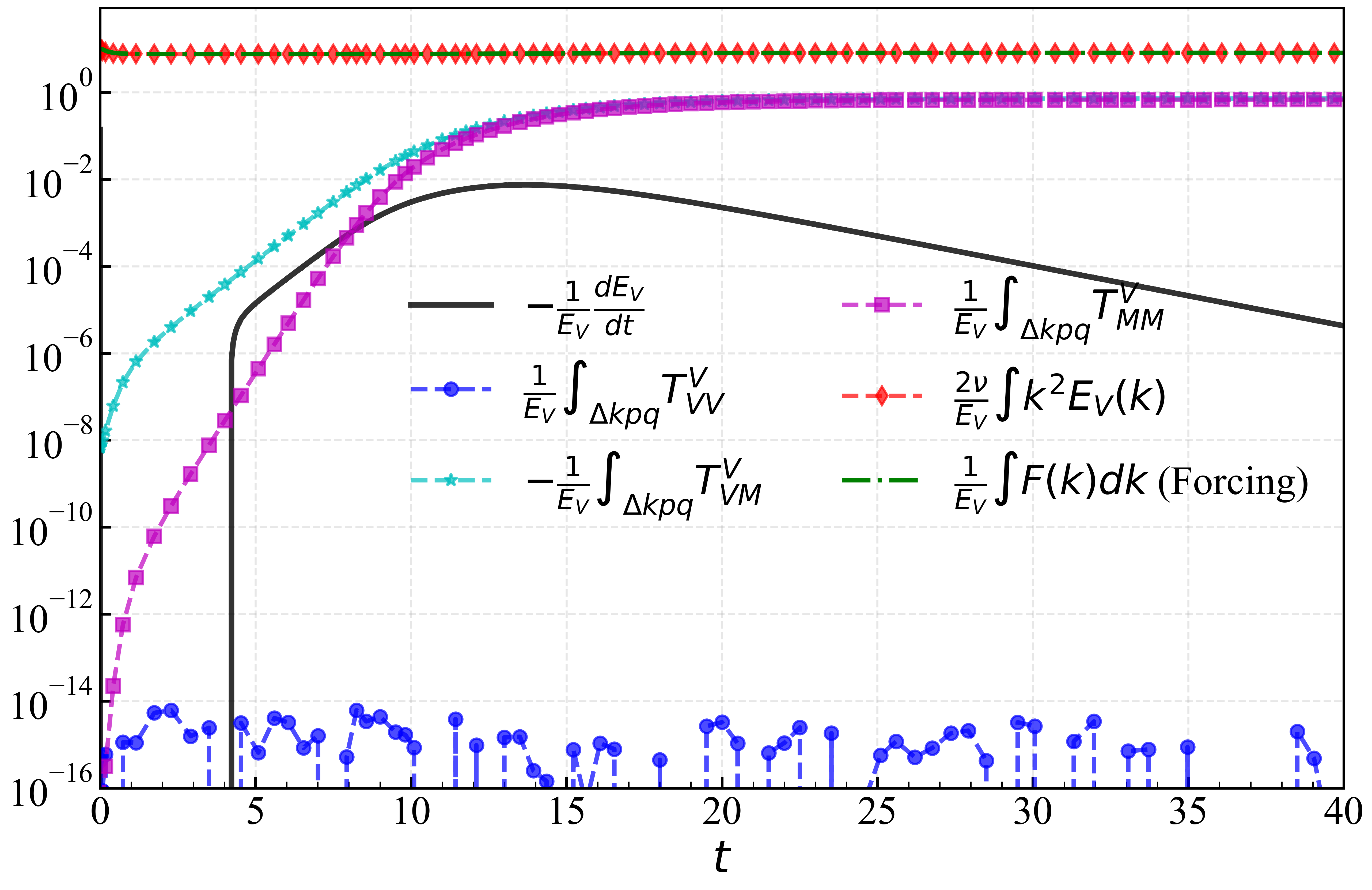}
    \caption{Term-by-term comparison of Eq.~\eqref{eq:cl1} integrated over $k$ and normalized by KE in the nonlinear `Kazantsev dynamo'. 
    Note that $\int_{\Delta kpq}\ \TVVM$ (cyan stars) is always negative. 
    This is balanced by $\int_{\Delta kpq} \TVMM$ (magenta squares). The green line represents the total energy injection by forcing, and the red line with diamond markers shows the total viscous dissipation. 
    The black line indicates the rate of change of the kinetic energy.}
    \label{fig:kinterms_vs_t_fullkaz}
\end{figure}

{Similarly, a term-by-term analysis of Eq.~\eqref{eq:cl1} is shown in }
Fig.~\ref{fig:kinterms_vs_t_fullkaz}{.} 
{The energy injection (green dash-dotted line), balances the viscous dissipation (red diamonds) for most of the time.
The interactions within the velocity only redistribute kinetic energy, thus $\int_{\Delta kpq} \TVVV$ shown in blue dots is zero to machine precision. The term $\int_{\Delta kpq} \TVVM$ (cyan stars) is the term conjugate to the stretching term $\TMVM$ and is always negative, {since the dynamo transfers KE to ME}. This causes the reduction in KE observed in Fig.~\ref{fig:Emag-kazfull} and shown by the black line in Fig.~\ref{fig:kinterms_vs_t_fullkaz}. Meanwhile, the interaction of magnetic fields given by $\int_{\Delta kpq} \TVMM$ (magenta squares) feeds the kinetic energy significantly in the nonlinear stage and a new steady state turbulence strength is achieved.}

\begin{figure}
    \centering
    \includegraphics[width=\linewidth]{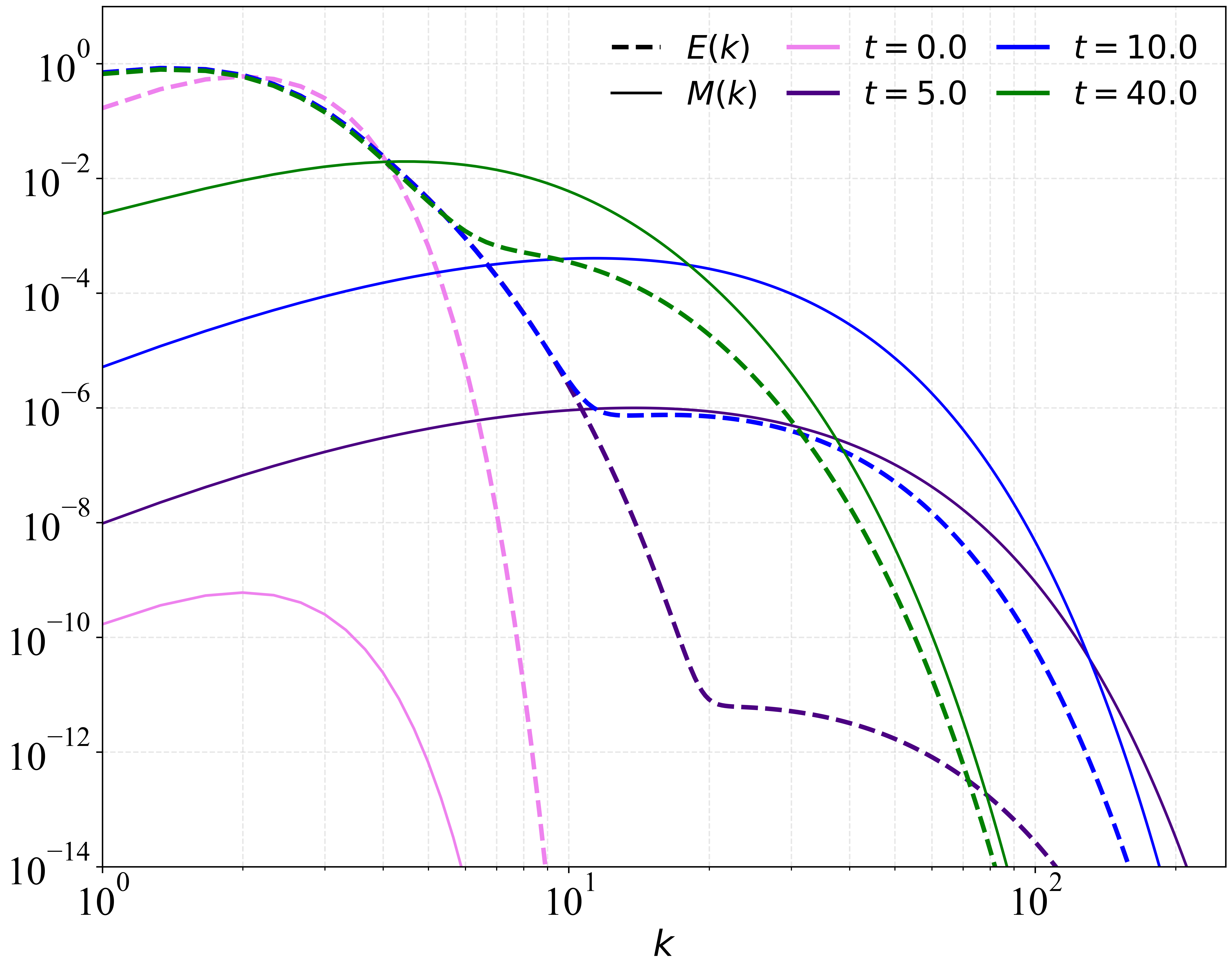}
    \caption{Evolution of KE (dashed) and ME (solid) spectra in the nonlinear `Kazantsev dynamo'. 
    }
    \label{fig:Especs-fullkaz}
\end{figure}

\begin{figure}
    \centering
    \includegraphics[width=\linewidth]{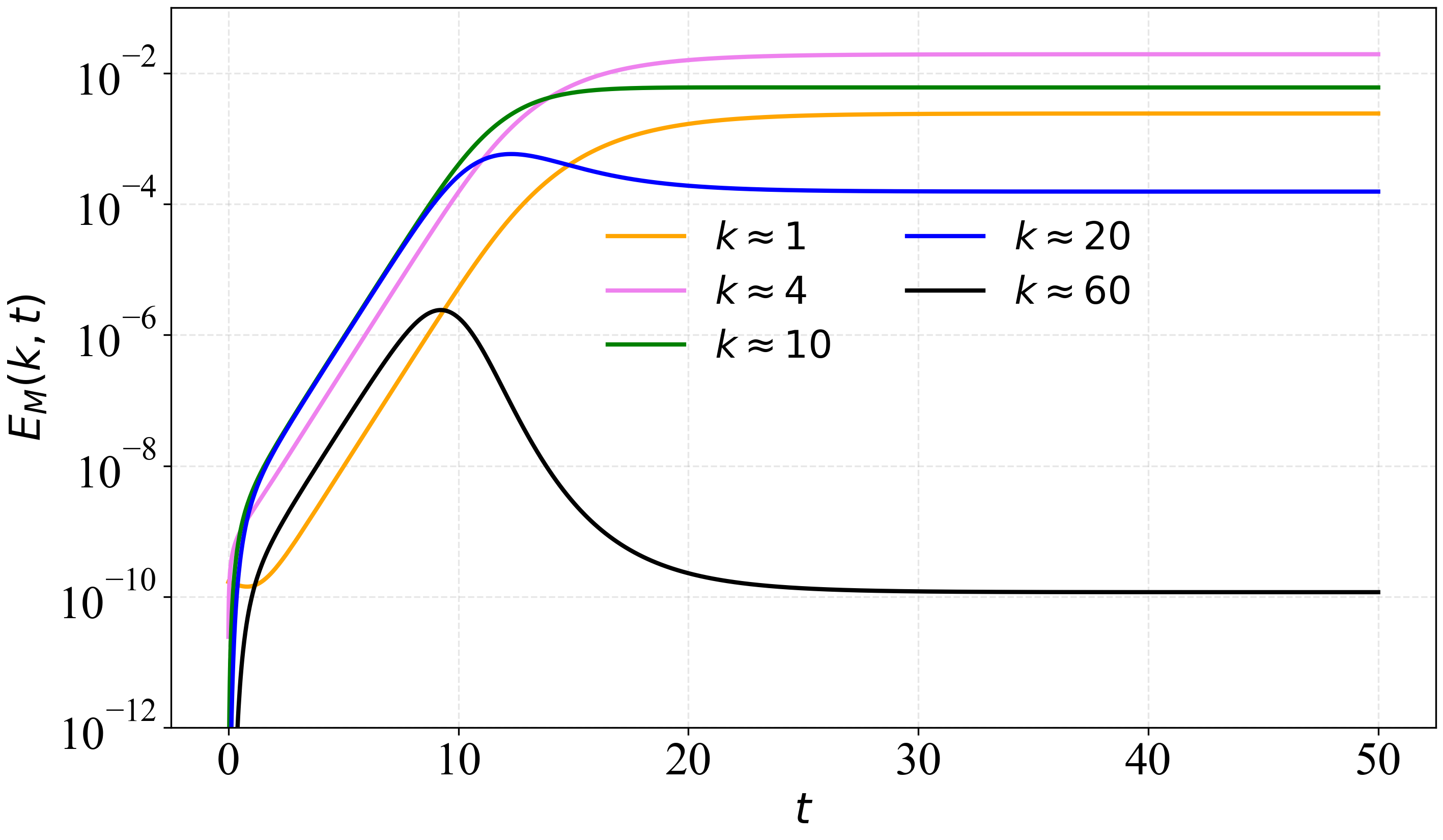}
    \caption{
    {Time evolution of different modes of ME spectrum in the case of nonlinear `Kazantsev dynamo'.}
    }
    \label{fig:Emag-modesfullkaz}
\end{figure}

The evolution {of KE and ME spectra} is shown in Fig.~\ref{fig:Especs-fullkaz}. 
{Following, the self similar growth in the kinematic stage (notice that in Fig.~\ref{fig:Emag-modesfullkaz}, different modes of ME grows with the same growth rate for sometime), the nonlinear interaction of magnetic field $\TMMM$}  shifts the peak of the ME spectrum  from the resistive scales {to larger scales} and suppresses small scale power. 
{In fact this reduction in $\kpeak$ from kinematic to nonlinear regime is by an order of magnitude.} 
The suppression in ME of the high $k$ modes is clear from Fig.~\ref{fig:Emag-modesfullkaz}. 

\begin{figure}
    \centering
    \includegraphics[width=\linewidth]{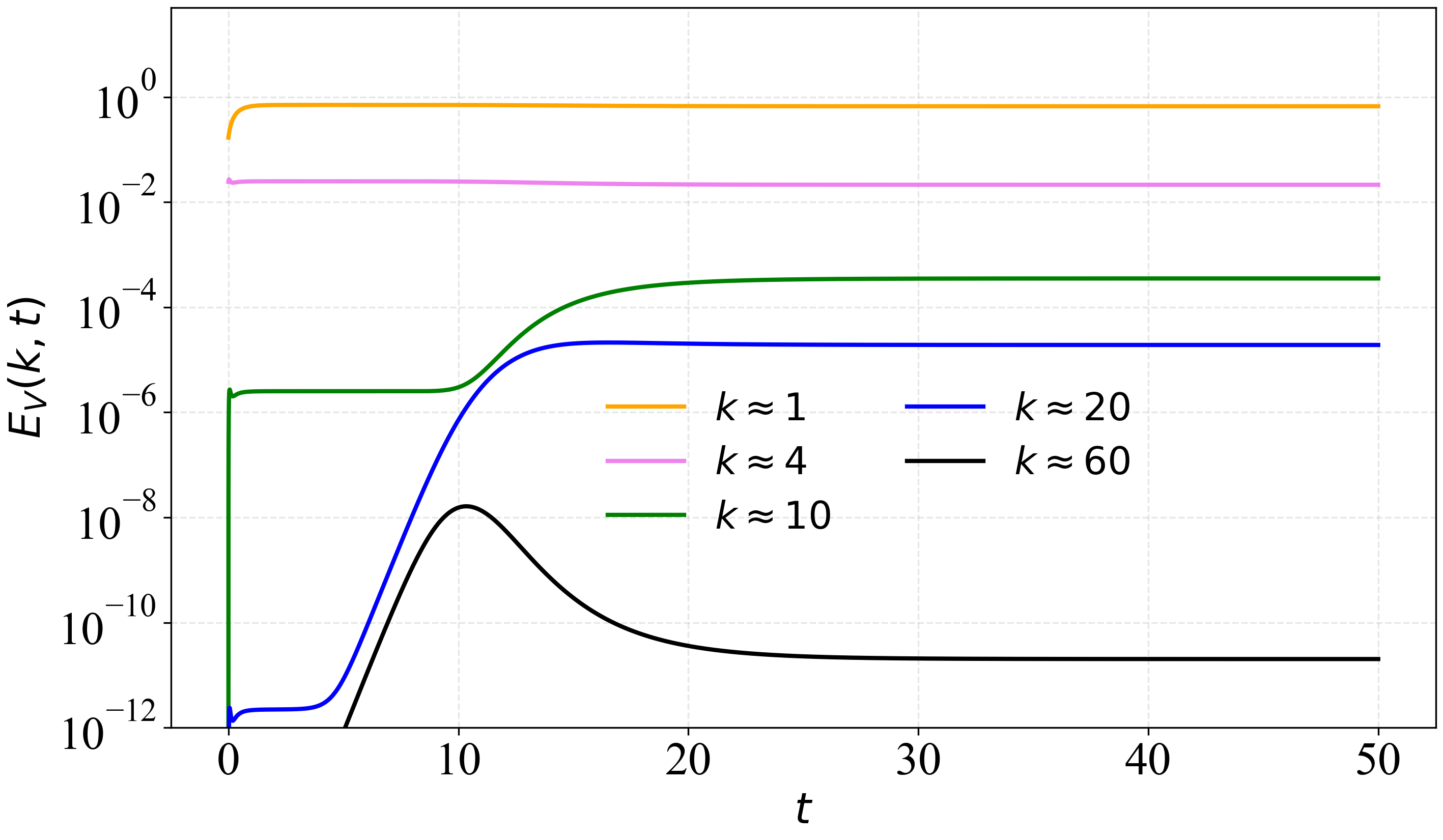}
    \caption{Time evolution of different modes of KE spectrum in the case of nonlinear `Kazantsev dynamo'.}
    \label{fig:Ekin-modesfullkaz}
\end{figure}

Because of the high $\Pm$, the ME exceeds the KE at small scales even during the kinematic stage. At this point the magnetic energy transfers to KE, opposite of the dynamo action {perhaps due to the nonlocal interactions between velocity and magnetic fields known as Alfv\'{e}nisation (see section 3 of \citet{PFL1976}).}
This feedback causes the KE spectrum to flatten at high wavenumbers.
The temporal evolution of the corresponding KE modes in Fig.~\ref{fig:Ekin-modesfullkaz} illustrates this as a secondary enhancement in their amplitude following their initial hydrodynamic steady state.
{While there is enhancement on small scale {kinetic energy}, {the} larger scales {experience} 
suppression by {the} Lorentz force, causing an overall reduction in the average velocity field amplitude. {Similar results are shown in the simulations by \cite{Brandenburg2019}}. 
It is important to note that the 
{Alfv\'{e}nisation} seems crucial in achieving equipartition between KE and ME at small scales.}
This observation has been possible as EDQNM simulation allows us to focus on small-scale physics with greater resolution.

{While Alfv\'{e}nisation brings the viscous and resistive scales closer, there is no scale by scale equipartition between KE and ME. Instead,} at saturation, the magnetic energy exceeds kinetic energy on small scales ($k\gtrsim 4$), similar to the results from DNS \cite{Haugen2004}. But the large scale ME is $\sim 100$ times smaller compared to the outer scale motions.

\begin{figure}
    \centering
    \includegraphics[width=\linewidth]{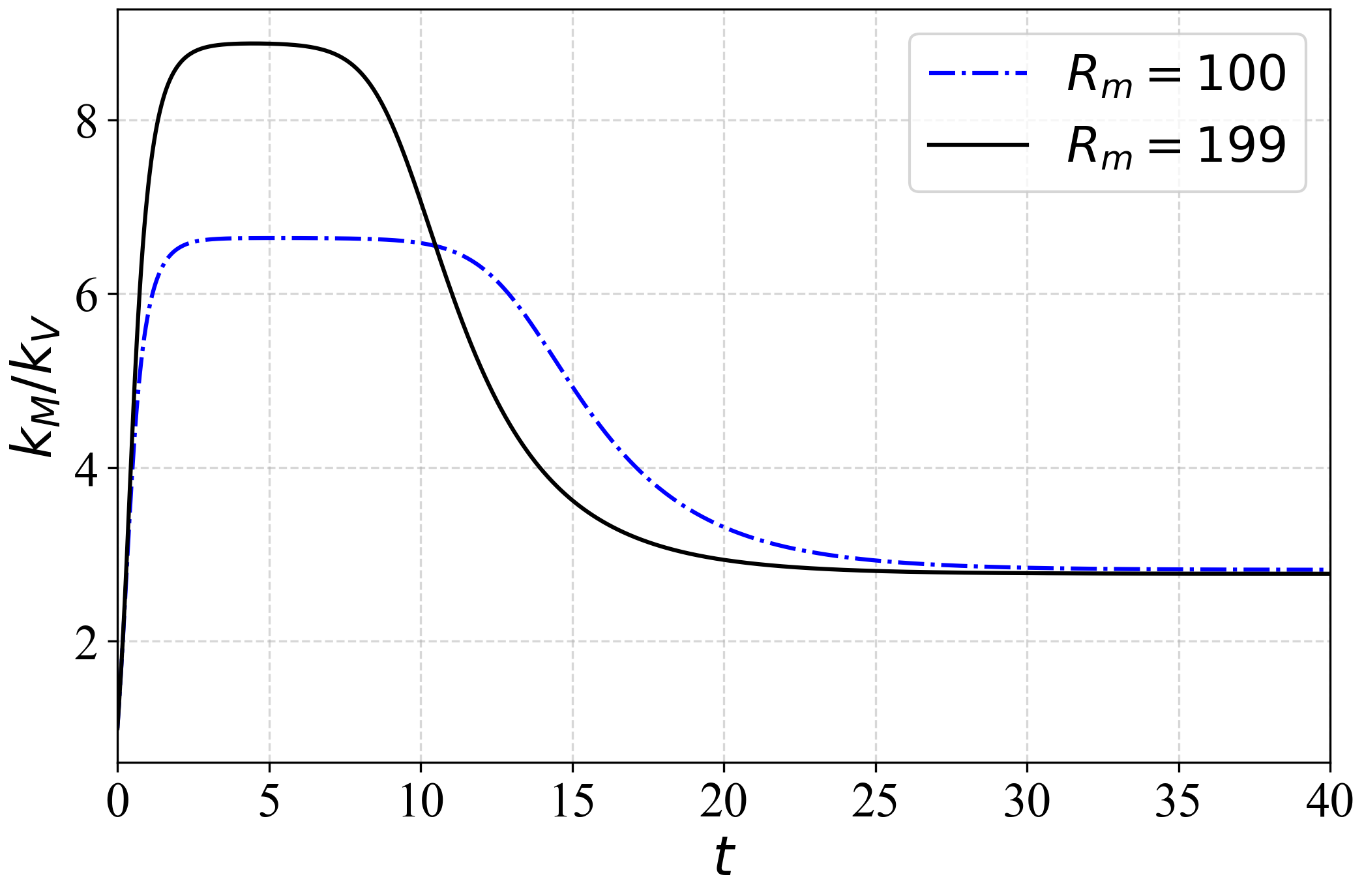}
    \caption{Comparing the evolution of the ratio of integral wavenumbers of ME and KE spectra $k_M / k_V$ in the nonlinear `Kazantsev dynamo' runs with $\Rm = 100$ and 199. Notice that at saturation they overlap to a common value.}
    \label{fig:kMkV_compare_fullkaz}
\end{figure}

A key result is the evolution of the ratio between integral wavenumber of ME spectra and KE spectra, defined as \cite{monin2013statistical}
\begin{align}\label{eq:int_k}
    \frac{k_M}{k_V} = \frac{\tfrac{1}{E_V}\int k^{-1} E(k) \, dk}{\tfrac{1}{E_M}\int k^{-1} M(k) \, dk}.
\end{align}
Figure~\ref{fig:kMkV_compare_fullkaz} shows that in the kinematic stage it is larger for larger $\Rm$, but at saturation it reaches an 
{asymptotic} value, independent of $\Rm$.
This suggests that nonlinear interactions effectively re-normalize the system toward an asymptotic state. 

\subsection{Effects of nonlinearity}
The nonlinear saturation of the dynamo can be understood by analyzing the transfer term $\TMMM$, which governs the evolution of magnetic energy in presence of back reaction
\begin{align}\label{eq:cl8a}
    I &= \tau \int_{\Delta_k} \dd p\ \dd q\ \TMMM = -\tau k^3\ M_k\ \int_{\Delta_k} \frac{\dd p\ \dd q}{pq}\ c_{kpq}\ M_q.
\end{align}
\textbf{Large scale limit ($k \ll q$):}
On larger scales this simplifies to (see appendix \ref{sec:tmmm})
\begin{align}\label{eq:cl8}
    I &= - \frac{4}{5}\tau\ k^4 M_k \left(\int \dd q \ \frac{M_q}{q^2}\right).
\end{align}
Thus the back reaction from magnetic field leads to a hyper diffusion of magnetic energy on larger scales with diffusivity proportional to magnetic energy. 

\begin{figure}
    \centering
    \includegraphics[width=0.48\textwidth]{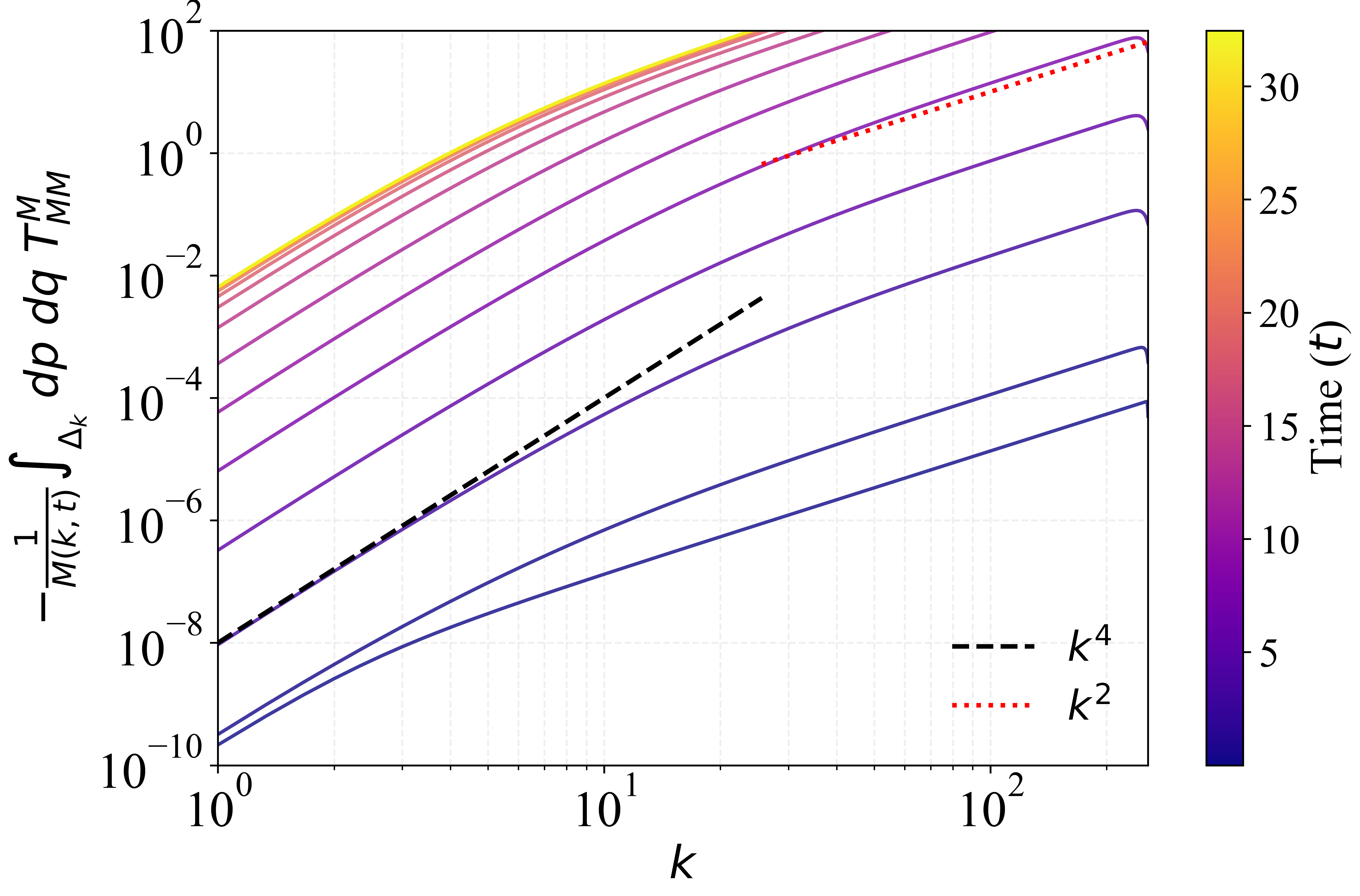}
    \caption{Negative of $\intDk \ \dd p\ \dd q\ \TMMM / M_k$ at different times (shown in the color bar) as a function of $k$. As theoretically expected, the large scale shows hyper-diffusive $k^4$ scaling (dashed line), while the smaller scales experience an additional $k^2$ diffusion (dotted line).}
    \label{fig:TMMM_k_kazfull}
\end{figure}

\textbf{Small scale limit ($k \gg q$):}
Similarly, on small scales the integral becomes (refer to appendix \ref{sec:tmmm})
\begin{align}\label{eq:cl9}
    I &= -\frac{4}{3}\tau\ k^2 M_k \left(\int M_q \dd q\right) 
      + \frac{4}{5}\tau\ M_k \left(\int \dd q\ q^2 M_q\right).
\end{align}
The first term is the turbulent dissipation of magnetic energy, similar to Eq.~\eqref{eq:cl7b} with the coefficient proportional to ME instead of KE. And the positive definite second term amplifies the magnetic spectra. But in this limit of $k \gg q$
\[
\int \dd q\ q^2 M_q < k^2 \int \dd q\ M_q, \implies I < -\frac{8}{15}\tau\ k^2 M_k \left(\int M_q \dd q\right). 
\]
Therefore, the nonlinear term is negative definite on small scales, implying that diffusion dominates over nonlinear stretching.
These limits we verify from the simulations of last section and shown in Fig.~\ref{fig:TMMM_k_kazfull}.
Therefore, nonlinearity introduces an effective diffusion that increases with magnetic energy, suppressing growth at small scales and leading to saturation.
{Interestingly, these results on the effect of nonlinear back reaction are qualitatively similar to those
obtained in the somewhat simpler model of \cite{Subramanian2003Hyperdiffusion}.}

To understand the effect of magnetic field on KE spectra we perform the integral on $\TVVM$ (see appendix \ref{sec:tvvm}), 
\begin{align}\label{eq:cl10}
    \tau\intDk \dd p\ \dd q\ \TVVM &= -\tau E_k\intDk \dd p\ \dd q\ \frac{p^2}{q}\ z(1-y^2) M_q \notag \\
    &= -\frac{4}{3}\tau k^2 E_k \int\dd q\ M_q.
\end{align}
Magnetic field interaction with velocity provides additional dissipation with a diffusion coefficient proportional to the magnetic energy.
Then it is clear that $\TVMM$ should cause the flattening of KE spectra. 
But it is hard to integrate out that and identify the sourcing term analytically.

\begin{figure}
    \centering
    \includegraphics[width=0.48\textwidth]{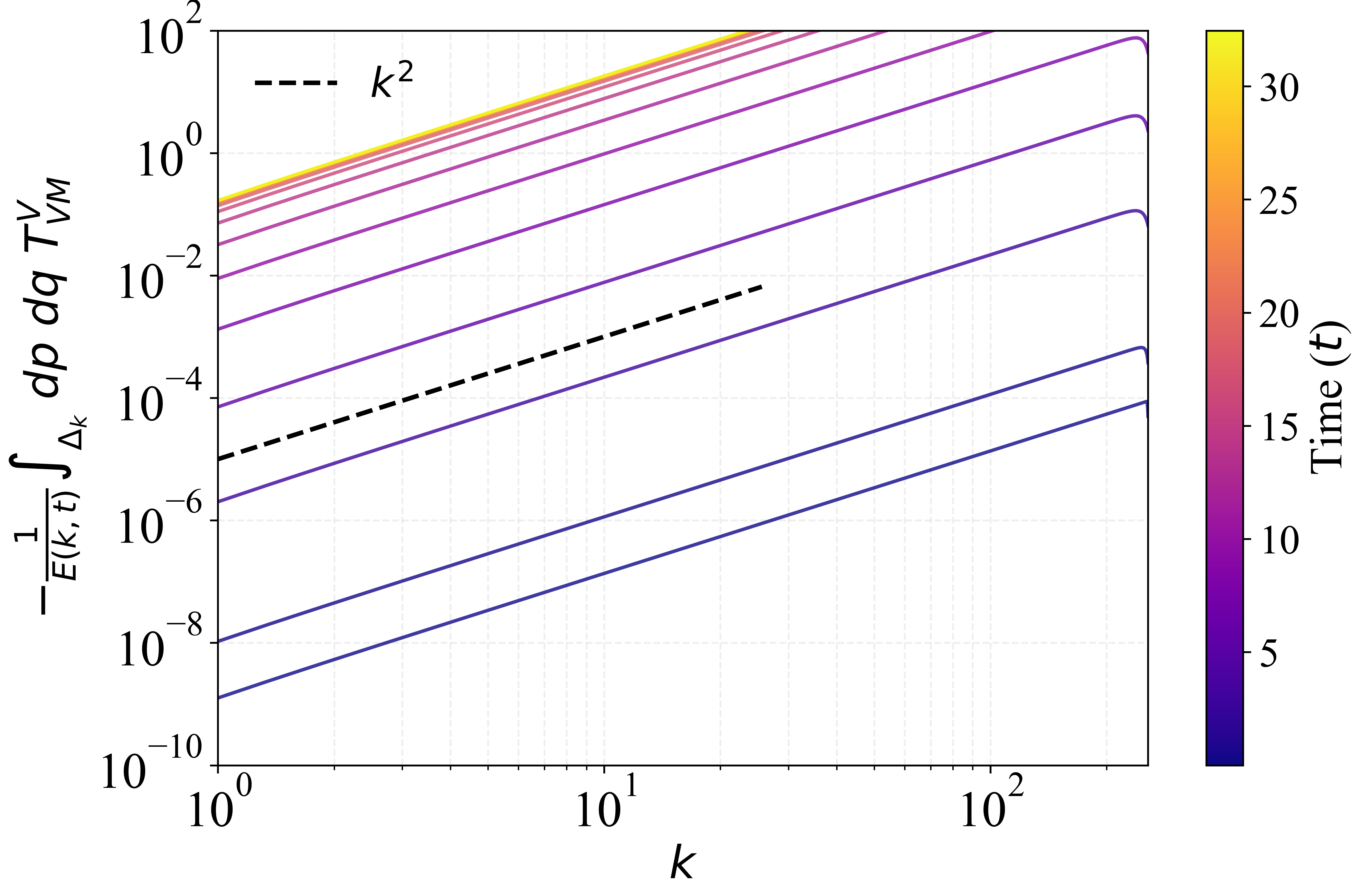}
    \caption{Negative of $\intDk \ \dd p\ \dd q\ \TVVM / E_k$ at different times (shown in the color bar) as a function of $k$. As expected from the theory, it is a diffusion term (see the $k^2$ scaling shown in dashed).}
    \label{fig:TVVM_k_kazfull}
\end{figure}

\section{Simulations - EDQNM}\label{sec:edqnm_sim}
We have shown analytically and numerically that the nonhelical EDQNM equations recover the  {Kazantsev} dynamo description 
{and its simple nonlinear extension} when $\thkpq = \tau$.
Here, we perform a suite of simulations of Eqs~\eqref{eq:cl1}~–~\eqref{eq:cl6} using the scale-dependent eddy damping  $\thkpq$, which is computationally less expensive than the case $\thkpq = \tau$ (see the discussion in Appendix \ref{num-setup}).
We explore the full $(\Rey, \Rm)$ parameter space accessible numerically, thereby covering a wide range of magnetic Prandtl numbers $\Pm = \Rm/\Rey$.
Unless stated otherwise, the forcing is applied around $k_f = 2$.

\subsection{Critical magnetic Reynolds number}
\begin{figure}
    \centering
    \includegraphics[width=\linewidth]{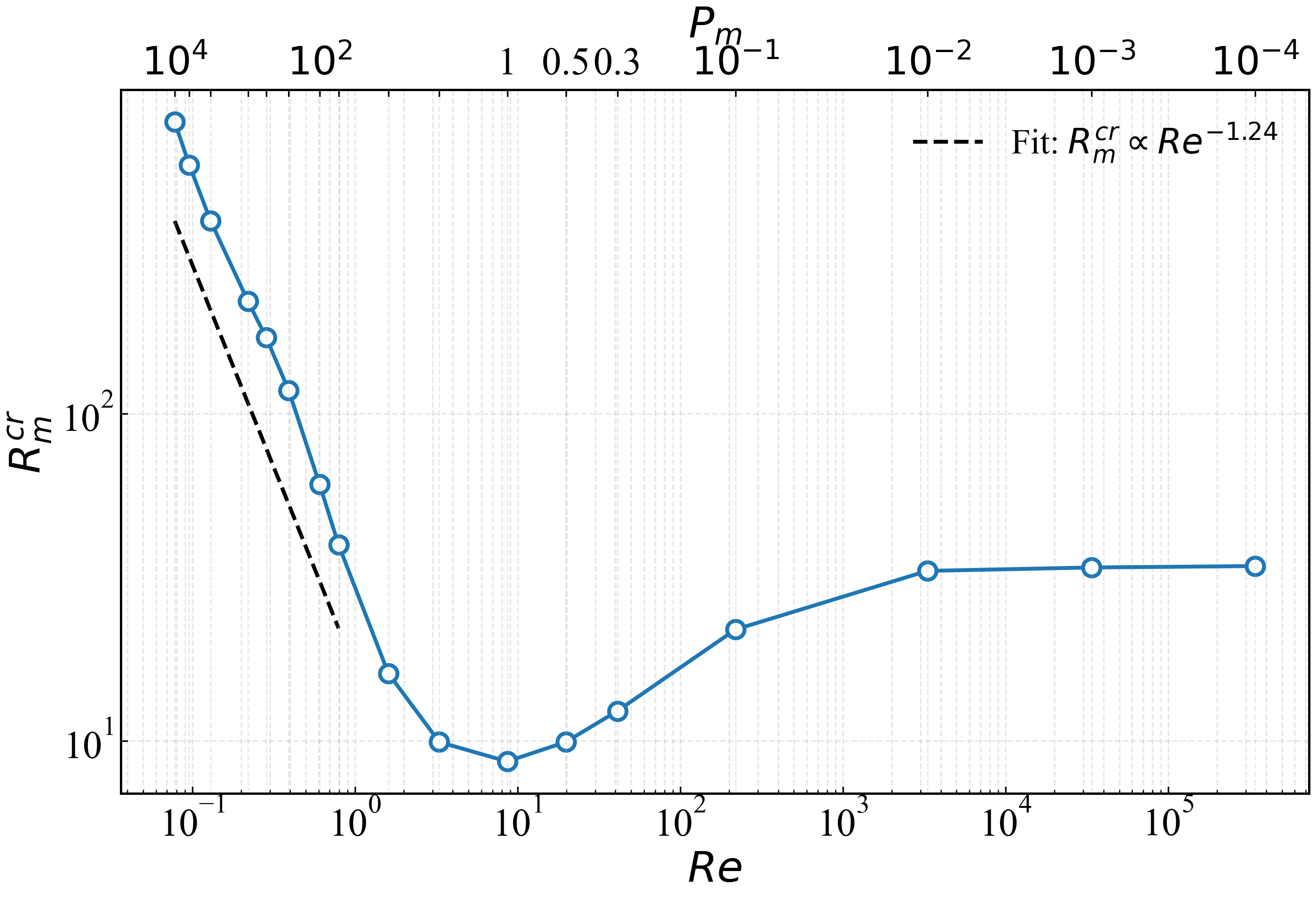}
    \caption{Critical magnetic Reynolds number ($\Rmcr$) as a function of $\Rey$ and corresponding $\Pm$.
    }
    \label{fig:Rmcr_Re_Pm}
\end{figure}

Magnetic field amplification occurs when $\Rm > \Rmcr$, such that stretching overcomes 
dissipation. We determine $\Rmcr$ numerically for a given $\Pm$ by {iteratively varying $\Rm$ and} identifying the point of zero {magnetic} growth rate. As shown in Fig.~\ref{fig:Rmcr_Re_Pm},
for $\Pm = 1$, $\Rmcr$ attains a minimum value of $\sim 10$, which lies on the lower end compared to earlier theoretical and numerical estimates \citep{subramanian1997dynamics, Haugen2004, Schekochihin_2007}. 
{$\Rmcr$ increases with decreasing kinetic Reynolds number approximately as $\Rmcr \propto \Rey^{-5/4}$ in the viscous regime ($\Rey\lesssim 10$). In the turbulent regime ($\Rey\gtrsim 10^4$), $\Rmcr$ approaches a constant value. In other words, at $\Pm > 1$, $\Rmcr$ increases as $\Rmcr \propto \Pm^{5/9}$ and for $\Pm < 1$, $\Rmcr$ rises and asymptotes to $\sim 35$ by $\Pm=0.01.$}

{Qualitatively, we attribute this to the changing separation between the resistive and viscous scales. For $\Pm>1$, the resistive scale $k_\eta \sim k_\nu \Pm^{1/2}$ moves progressively away from the viscous scale $k_\nu$ as $\Pm$ increases. As a result, the dynamo eigenfunction must extend over a broader range of scales, making the onset of growth increasingly difficult and leading to a larger $\Rmcr$. In contrast, when $\Pm<1$, magnetic amplification is controlled primarily by inertial-range eddies near $k_\eta$. As $\Pm$ decreases, the separation between $k_\eta$ and $k_\nu$ increases, but the stretching dynamics around $k_\eta$ remain largely unchanged. Thus the eigenfunction becomes localized within the inertial range and loses sensitivity to the viscous cutoff, causing $\Rmcr$ to approach an asymptotic value independent of $\Pm$.}

\subsection{Growth rate}
\begin{figure}
    \centering
    \includegraphics[width=\linewidth]{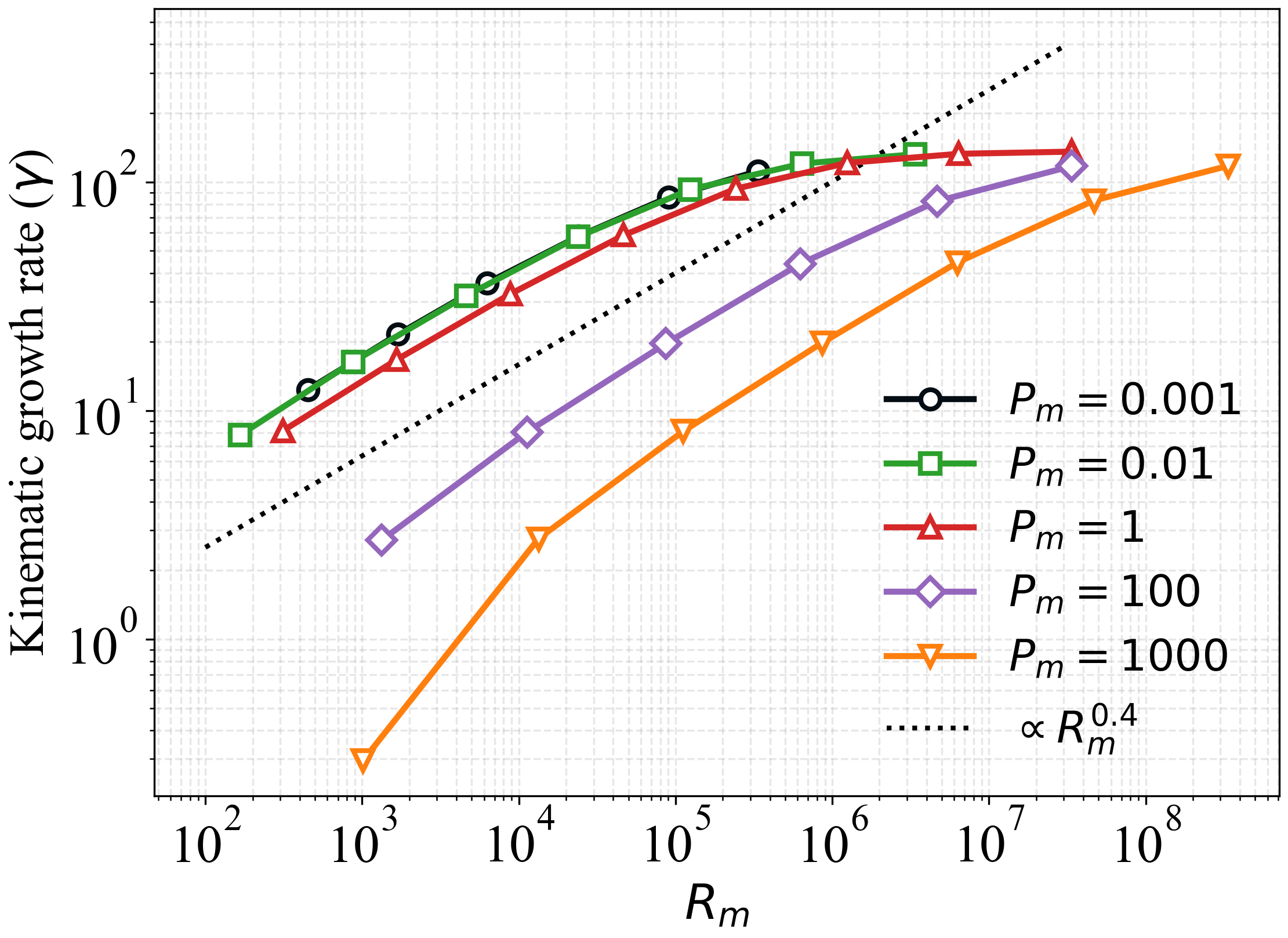}
    \caption{Growth rate in the kinematic stage ($\gamma$) as a function of $\Rm$ for different $\Pm$ values. The dotted line gives a reasonable scaling $\gamma \propto \Rm^{0.4}$ prior to the asymptotic regime.
    }
    \label{fig:gr_asym_pm}
\end{figure}

The kinematic growth rate of the magnetic energy ($\gamma$) is conventionally obtained from an exponential fit to the temporal evolution of the magnetic energy. However, at large $\Rey$ and $\Rm$, the dynamo growth becomes extremely rapid, rendering such fits unreliable. {Thus,} instead, we extract the maximum value of $\dd \ln E_M / \dd t$ as a proxy for $\gamma$. The resulting growth rates as a function of $\Rm$ for various $\Pm$ are presented in Fig.~\ref{fig:gr_asym_pm}.

The orange inverted triangles and blue diamonds denote the large magnetic Prandtl number cases, $\Pm = 10^2$ and $10^3$, respectively. For a fixed $\Rm$, $\gamma$ is higher at smaller $\Pm$ (i.e., higher $\Rey$). Conversely, the growth rates are nearly identical when evaluated at similar $\Rey$ across different $\Pm > 1$ regimes. This occurs because, at large $\Pm$, the viscous-scale eddies are the fastest and dominate the magnetic field stretching, making the growth rate highly sensitive to $\Rey$. In contrast, for $\Pm \le 1$ shown by red triangles ($\Pm = 1$), green squares ($\Pm = 10^{-2}$) and black circles ($\Pm = 10^{-3})$, the resistive-scale eddies become the most effective at stretching. In this regime, the growth rate depends strongly on $\Rm$ 
{(at values lower than the asymptotic limit} 
{discussed below}),
and as expected, the $\Pm = 10^{-2}$ and $\Pm = 10^{-3}$ cases exhibit very similar $\gamma$ for a given $\Rm$.

At a fixed $\Pm$, increasing $\Rm$ eventually yields an asymptotic growth rate ($\Gasym$), a hallmark of fast dynamo. Importantly, $\Gasym$ approach a 
$\Pm$-independent value at a sufficiently high $\Rey = \Reasym \sim 10^6$ for $\Pm > 1$ and $\Rm = \Rmasym \sim 10^6$ for $\Pm \leq 1$ where the system is extremely turbulent. 
While \cite{Haugen2004} 
showed indications of this asymptote, they were unable to probe the requisite high $\Rey$ and $\Rm$ due to computational constraints. To the best of our knowledge, this is the first time such a clear asymptotic state has been established across a wide range of $\Pm$, albeit within a closure formalism.

Prior to reaching the asymptotic plateau, the growth rate follows a distinct power-law scaling of approximately $\gamma \propto \Rm^{0.4}$ (indicated by the dotted reference line). For classical Kolmogorov turbulence, one expects $\gamma_{\text{kol}} \propto \Rm^{1/2}$ when $\Pm \leq 1$, otherwise $\gamma_{\text{kol}} \propto \Rey^{1/2}$. The deviation from this scaling likely arises because the hydrodynamic kinetic energy spectrum is still developing during the kinematic dynamo stage, rather than being fully established and stationary.

\subsection{Saturation regime}
At the end of the kinematic stage, strong back reaction from the Lorentz force on the flow prevents further stretching of the magnetic field, leading to the saturation of the magnetic energy. 
For a constant damping time ($\thkpq = \tau$), we previously showed that $\TMMM$ leads to hyperdiffusion at large scales and enhanced diffusion at small scales within the magnetic energy spectrum. Concurrently, it induces enhanced dissipation {in the kinetic energy spectra} via the back reaction on the velocity field. Together, these mechanisms govern the physics of saturation 
{in our simulations}. 

\begin{figure}
    \centering
    \includegraphics[width=\linewidth]{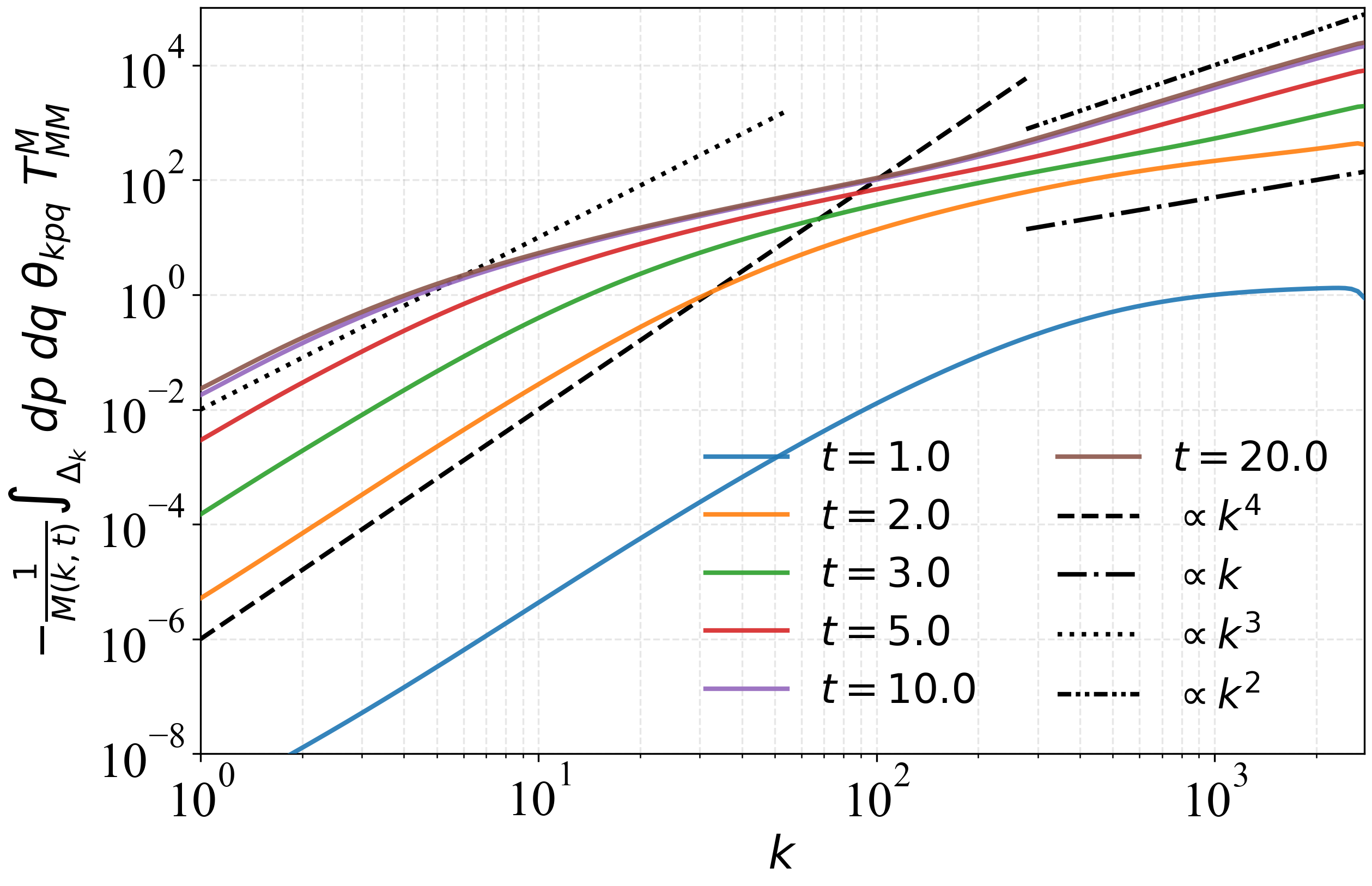}
    \caption{Evolution of the normalized magnetic transfer term
$-\intDk \dd p\, \dd q\ \theta_{kpq}\, \TMMM\, / M_k$ as a function of wavenumber at various simulation times.
The reference scalings {are given for} $\propto k^4$ (dashed), $\propto k^3$ (dotted), $\propto k^2$ (dash-dot-dotted) and $\propto k$ (dash-dotted).
}
    \label{fig:edqnm_tmmm}
\end{figure}

\begin{figure}
    \centering
    \includegraphics[width=\linewidth]{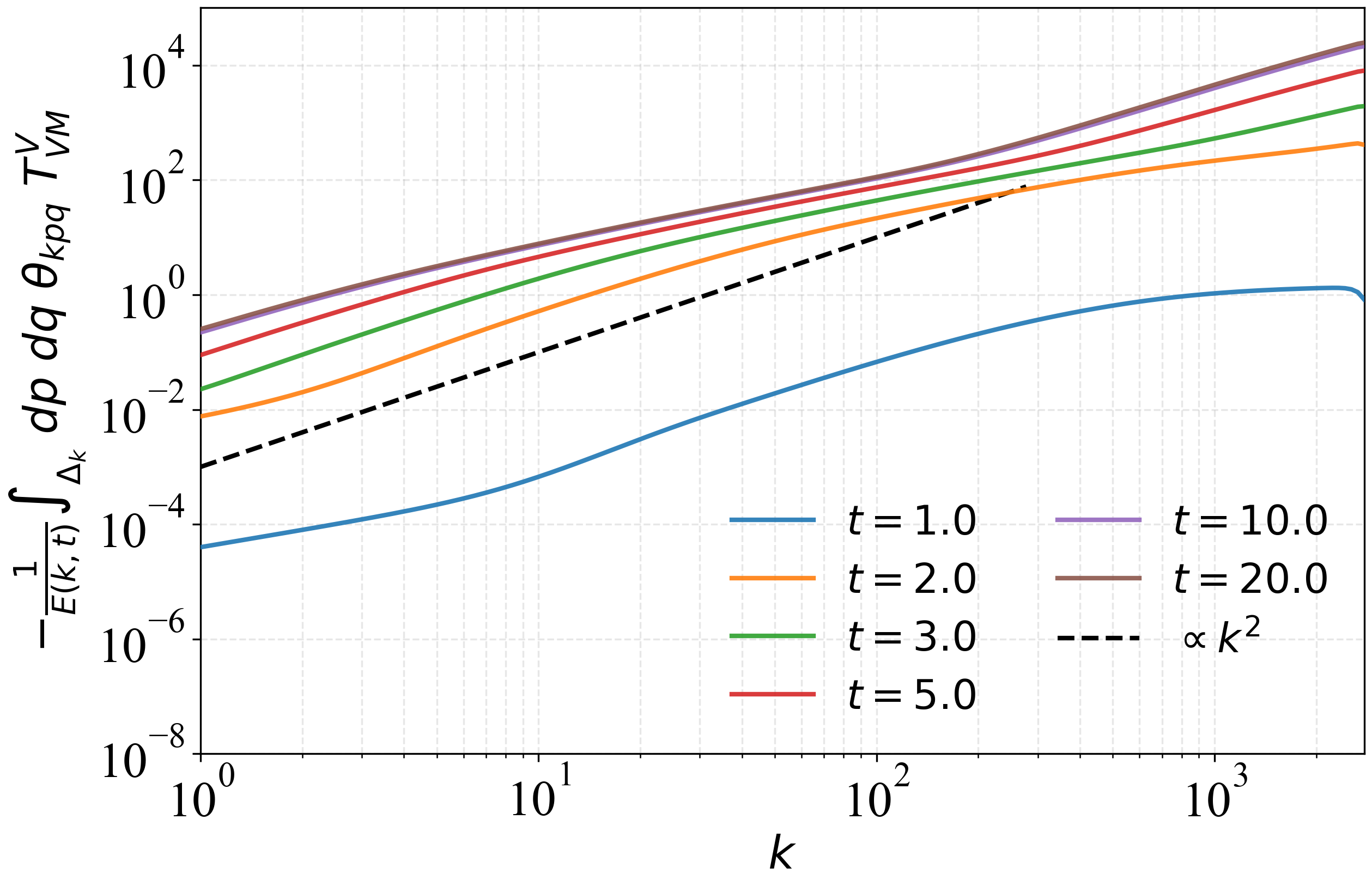}
    \caption{Similar to Fig.~\ref{fig:edqnm_tmmm}, but showing
$-\intDk \dd p\, \dd q\ \theta_{kpq}\, \TVVM\, / E_k$ as a function of wavenumber at various simulation times.
The dashed lines gives the scaling $\propto k^2$ and indicate diffusion.
}
    \label{fig:edqnm_tvvm}
\end{figure}

However, with scale-dependent damping, this behavior is significantly modified. As demonstrated in Fig.~\ref{fig:edqnm_tmmm}, {in the kinematic stage} the normalized integral $\intDk \dd p\, \dd q\, \thkpq \TMMM \propto -kM_k$ at small scales (dash - dotted line),
rather than standard diffusion (which scales as $-k^2 M_k$). Meanwhile, the hyperdiffusive scaling ($\propto k^4$) at large scales is preserved (dashed line). 
{At saturation, the large scales shows a scaling close to $ \propto k^3$ (dotted line) and at small scales $\propto k^2$ is observed.}
Similarly, the evolution of the normalized $\TVVM$ term shown in Fig.~\ref{fig:edqnm_tvvm} reveals standard diffusion ($\propto k^2$) at large and intermediate scales (dashed line), but transitions to only a sub-diffusive suppression of the kinetic energy spectrum at {relevant} small scales. Thus, the saturation mechanism depends crucially on the functional form of $\thkpq$. 

\begin{figure}
    \centering
    \includegraphics[width=\linewidth]{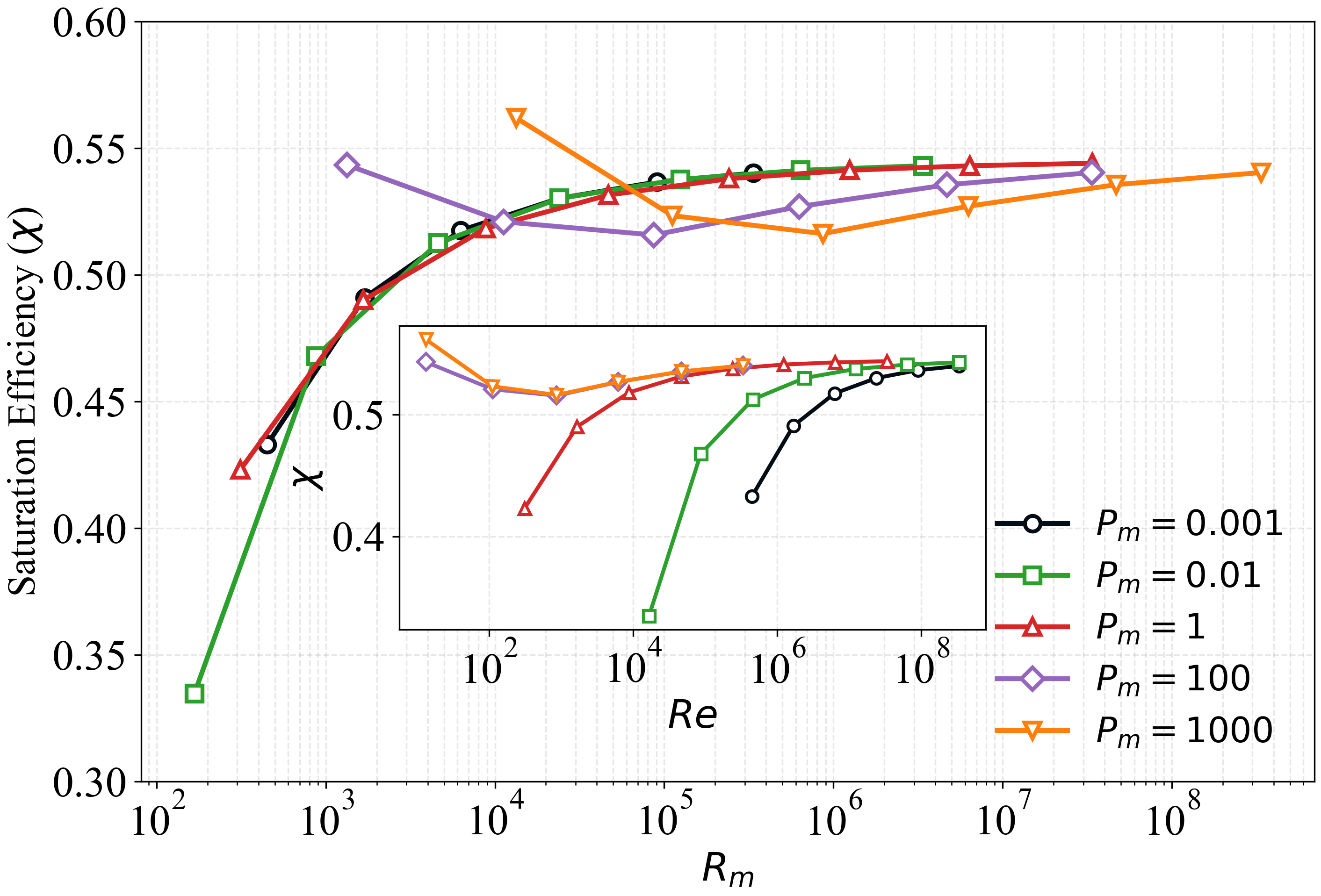}
    \caption{Saturation efficiency $\chi = \bsat^2 / \usat^2$ as a function of $\Rm$ for different $\Pm$ values. Similarly as a function of $\Rey$ is given in the inset.
    }
    \label{fig:chi_asym}
\end{figure}

We quantify the saturation level using the ratio of magnetic to kinetic energy at saturation, $\chi = \bsat^2 / \usat^2$, shown in Fig.~\ref{fig:chi_asym}. In $\Pm > 1$ {simulations}, $\chi$ {values lie on top of each other when presented as a function of $\Rey$ (see inset).} {Starting from} $\chi = 0.57$ {in the viscous case, it} changes slightly across the range {in $\Rm$ and $\Rey$}. 
Conversely, for $\Pm \le 1$, $\chi$ {values} overlap one another {when presented against $\Rm$ and increases with increasing $\Rm$ before reaching an asymptotic value.} 
Importantly, {the parameters for the asymptote are once again} $\Reasym \sim 10^6$ for $\Pm > 1$ and at $\Rmasym \sim 10^6$ for $\Pm \leq 1$. {And} $\chi$ asymptotes to approximately 55\% of the energy in turbulent motions {for all $\Pm$}. 

{To our knowledge, this asymptotic behaviour in saturation has not been demonstrated previously. Earlier discussions of small-scale dynamo saturation have largely focused on two limiting pictures: one in which nonlinearity modifies the effective transport coefficients, driving the system toward a marginally stable state, and another in which the saturated state retains an explicit dependence on $\Rm$. In contrast, the results here indicate the emergence of an $\Rm$-independent 
{(and $\Rey$-independent)} asymptotic regime once $\Rm$ {(and $\Rey$)} exceeds a sufficiently large threshold. 
This is physically plausible, and is analogous to the onset of Reynolds-number-independent behaviour in hydrodynamic turbulence at sufficiently high $\Rey$ \cite{Frisch1995}.}

\begin{figure}
    \centering
    \includegraphics[width=\linewidth]{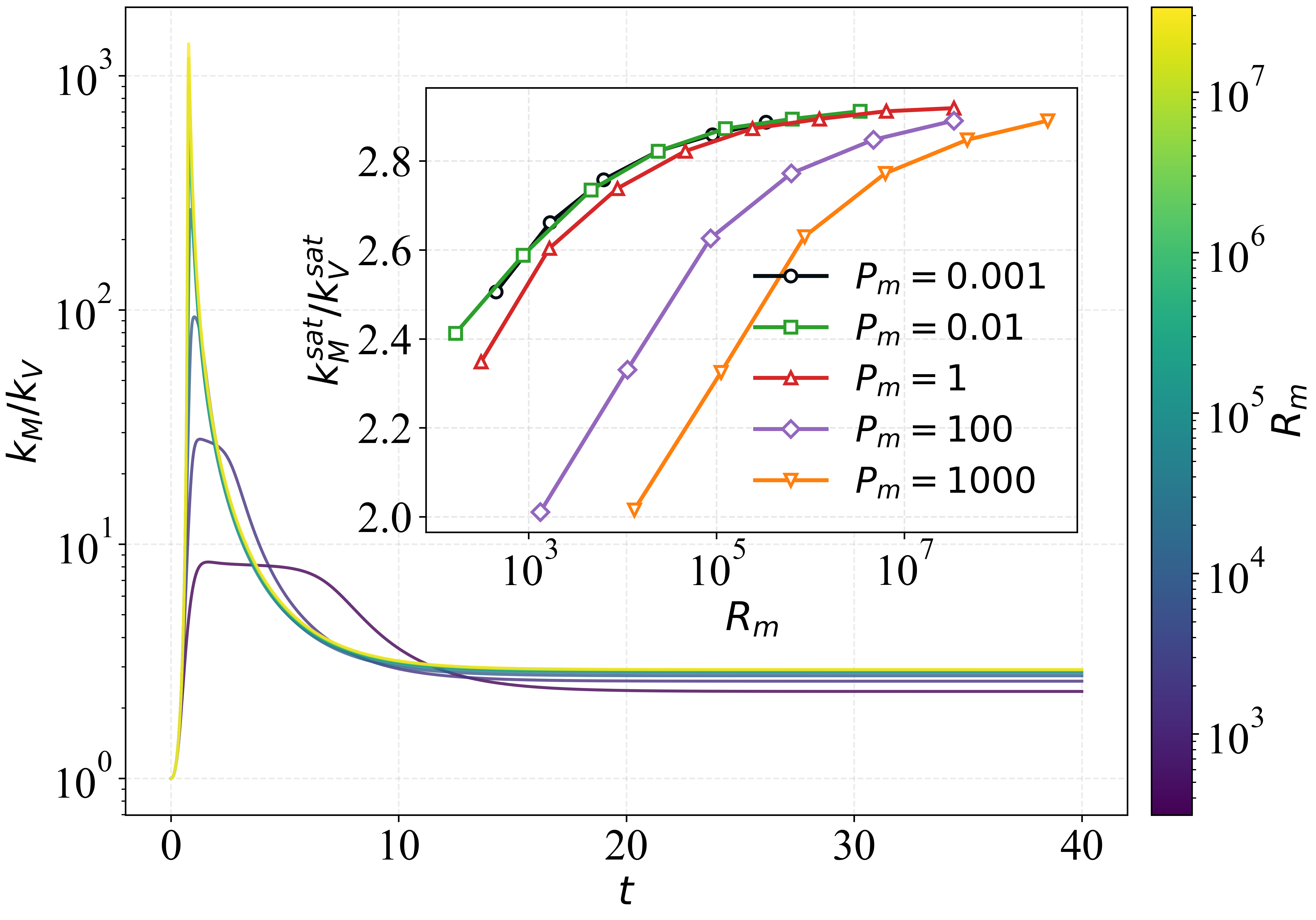}
    \caption{Evolution of the normalized magnetic integral scale, $k_M / k_V$. The main plot displays the time evolution for the $\Pm = 1$ case across various $\Rm$, given in the colorbar.  The inset shows the saturated ratio, $k_M^{\text{sat}} / k_V^{\text{sat}}$, as a function of $\Rm$ for different $\Pm$. All cases approach a universal asymptote of $k_M^{\text{sat}} / k_V^{\text{sat}} \approx 3$ at sufficiently high $\Rm$.}
    \label{fig:kMbykV_kf2}
\end{figure}

The temporal evolution of the magnetic integral wavenumber, $k_M$, normalized by the kinetic integral wavenumber, $k_V$, defined in Eq.~\eqref{eq:int_k} is shown in Fig.~\ref{fig:kMbykV_kf2}.
The main plot displays the evolution of $k_M / k_V$ for various  $\Rm$ at $\Pm = 1$. During the kinematic stage, the magnetic energy spectrum peaks at increasingly smaller scales as $\Rm$ increases, reflected by the sharp initial spike in the ratio. Remarkably, as the dynamo saturates, $k_M / k_V$ relaxes and converges to values between $2 - 3$ across all $\Rm$. 

The inset of Fig.~\ref{fig:kMbykV_kf2} focuses on this saturated ratio, $k_M^{\text{sat}} / k_V^{\text{sat}}$, as a function of $\Rm$ for different $\Pm$. For all $\Pm$ cases, the saturated scale separation increases 
with $\Rm$ and eventually asymptotes to $k_M^{\text{sat}} / k_V^{\text{sat}} \approx 3$ when reaching the highly turbulent asymptotic regime of $\Rm = \Rmasym \sim 10^6$ for $\Pm \leq 1$ and $\Rey = \Reasym \sim 10^6$ for $\Pm > 1$ consistent with estimates from Figs~\ref{fig:chi_asym} and \ref{fig:gr_asym_pm}. Furthermore, similar to $\chi$ for $\Pm \le 1$, $k_M^{\text{sat}} / k_V^{\text{sat}}$ the curves nearly overlap when plotted against $\Rm$, and for $\Pm > 1$ they overlap when the abscissa is $\Rey$. 

{The coherence scale at saturation is not independent of the saturation level \(\chi\) in our effective one-dimensional model, since the saturated magnetic energy is tied to the scale at which the magnetic spectrum peaks. Previous
models have estimated the saturated magnetic scale as
${k_M^{\rm sat}}/{k_V^{\rm sat}} \sim (\Rm^{\rm sat})^{-1/2}$,
with \(\Rm^{\rm sat}\) identified either with the imposed value \(\Rm\), implying
little change from the kinematic regime, or with \(\Rmcr\), corresponding to
saturation by renormalization to marginality. Our results suggest instead that the relevant effective magnetic Reynolds number may be an asymptotic value, ($\Rm^{\rm asym}$), associated with the nonlinear saturated regime. Nonetheless, it remains unclear whether simple stretching–diffusion balance arguments provide an adequate description of nonlinear saturation, especially in light of the asymptotic regime revealed by the EDQNM analysis.}

\subsection{Comparison of spectra}
Finally, we compare the evolution of kinetic and magnetic energy spectra for $\Pm \gg 1$ and $\Pm\ll 1$. 
{It is useful to notice that the large-scale magnetic field could acts as 
a {nearly} uniform background for the small-scale fluctuations, exciting Alfv\'{e}n waves that couple the {small-scale} velocity and magnetic fields and drive a
transfer of energy between them. This energy exchange is known as \textit{Alfv\'{e}nisation} (see Eqs~(3.1) and (3.2) of \citet{PFL1976} and discussions there). Consequently, if $M_k \neq E_k$ energy flows from the dominant to the subdominant field {at a rate proportional to the strength of the large-scale field}.}

\begin{figure}[htbp]
    \centering
    \includegraphics[width=\linewidth]{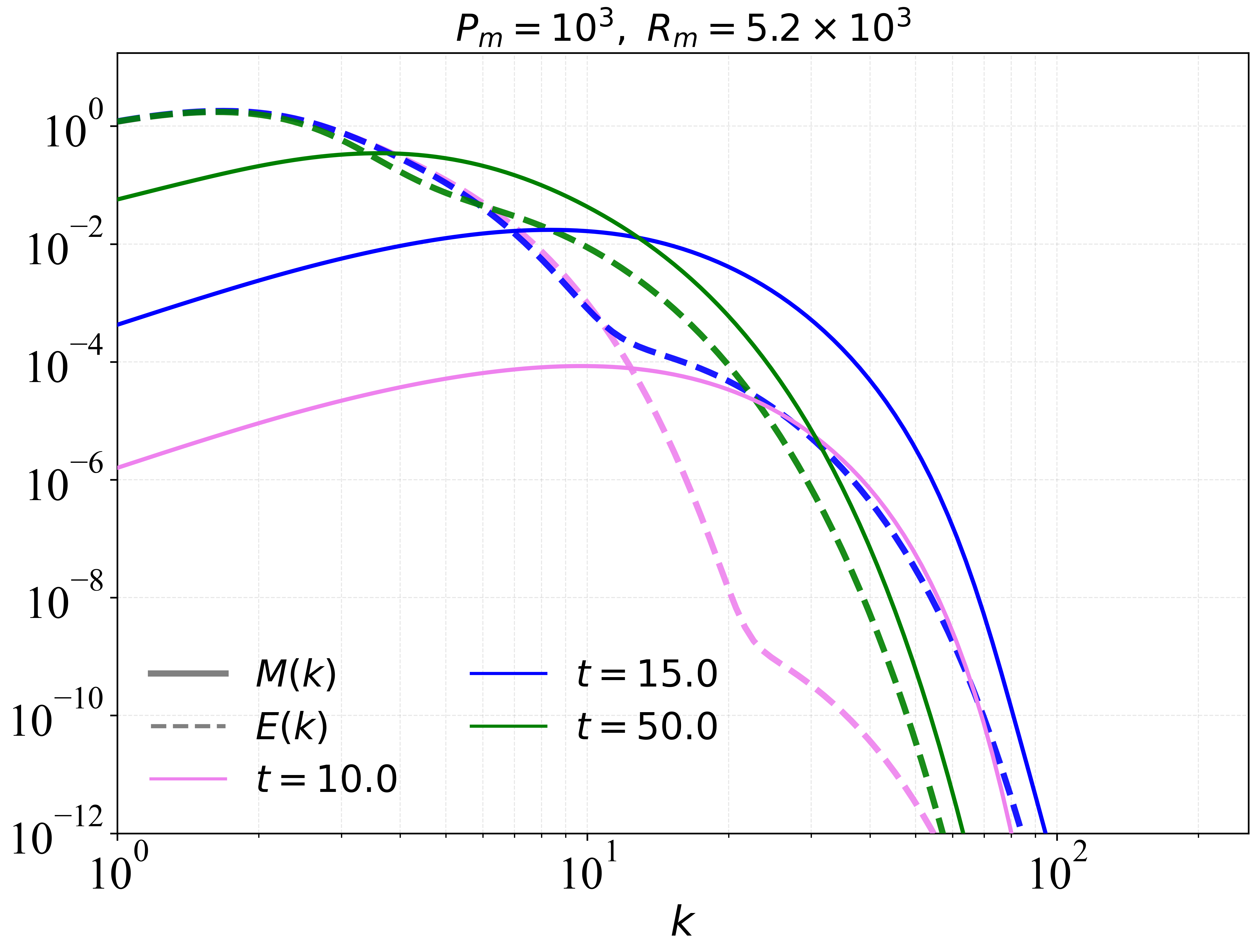}
    \caption{Evolution of the kinetic (dashed) and magnetic (solid) energy spectra for a $\Pm=10^3$ run with $\Rm \sim 5.2\times 10^3$ at various times. Notice the flattening in KE spectra due to the feedback from ME.}
    \label{fig:pm1e3_spec_evol}
\end{figure}

In the {case of $\Pm \gg 1$}, $k_\eta > k_\nu$, and {as shown in Fig.~\ref{fig:pm1e3_spec_evol} for the viscous and supercritical flow,} the magnetic energy (solid line) exceeds the kinetic energy (dashed) at small scales 
{and} the KE spectrum {flattens} {at these scales, {likely} due to} Alfv\`{e}nisation. 
Note that this modification is weaker than in the $\thkpq = \const$ model due to the stronger scale-dependent damping at smaller scales. 
{And it is worth noticing that, at saturation ME shows super equipartition with KE at wavenumbers $k > k_M$.} 

\begin{figure}
    \centering
    \includegraphics[width=\linewidth]{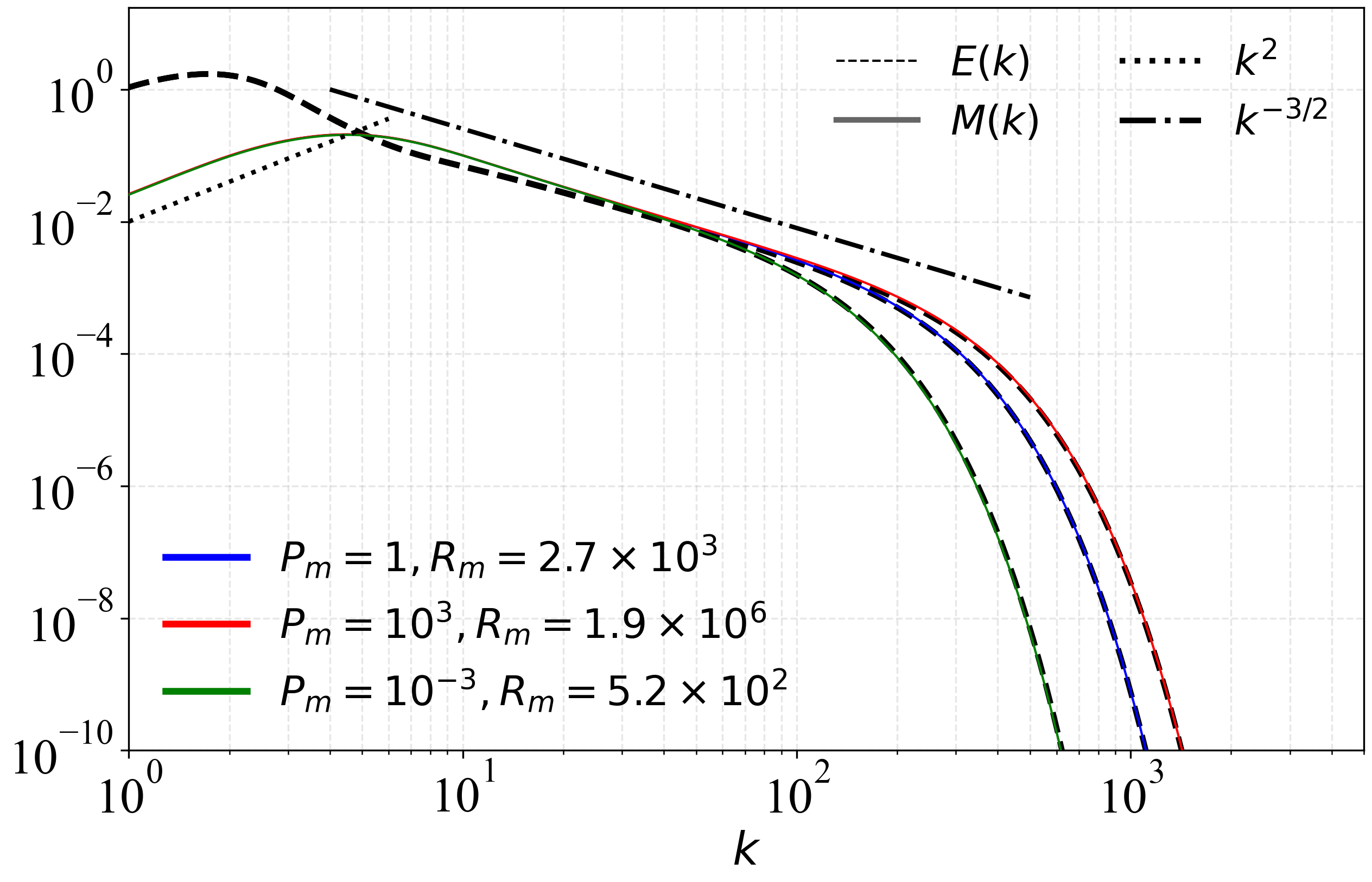}
    \caption{Comparison of the KE (dashed) and ME (solid) spectra at saturation for three different sets of parameters: $(\Pm, \Rey) = (10^{-3}, 5.2\times 10^5)$ in green, $(1,\, 2.7\times 10^3)$ in blue, and $(10^3,\, 1.9\times 10^3)$ in red. The $k^2$ (dotted) and $k^{-3/2}$ (dash-dotted) lines are shown for reference. 
    }
    \label{fig:pm_sat_compare}
\end{figure}

Figure~\ref{fig:pm_sat_compare} compares the final saturated states of the $\Pm = 1$, $\Pm > 1$, and $\Pm < 1$ cases, which are turbulent with high $\Rey$. The ME spectrum follows $M(k) \propto k^2$ (dotted line) at large scales and both $M(k)$ and $E(k)$ roughly follow a $k^{-3/2}$ scaling (dash-dotted lines) in the inertial range. {Although the resistive and viscous cutoff wavenumbers increase with $\Rm$,} the KE and ME spectra overlap almost perfectly {in the inertial range.}  This overlap implies that the effective viscous and resistive scales become nearly identical, causing the saturated turbulence to 
effectively resemble a $\Pm = 1$ system regardless of the initial Prandtl number. 
{These could be due to  Alfv\`{e}nisation transferring energy between magnetic and kinetic energy spectra to eventually balance them}.

So we conclude that when the system is highly turbulent, it goes to an effective $\Pm=1$ regime irrespective of the initial $\Pm$. This could be the reason for various asymptotic behaviors we found earlier.

\section{Conclusions and Discussions}\label{sec:discussion}
In this work, we use the eddy-damped quasi-normal Markovian (EDQNM) closure to study the growth and nonlinear saturation of the small-scale dynamo over a broad range of $\Rey$, $\Rm$, and $\Pm$. The EDQNM framework provides a computationally efficient way to probe parameter regimes that are difficult to access with 3-D direct numerical simulations, while still retaining the spectral transfer terms that encode velocity--magnetic and magnetic--magnetic interactions. This allows us to examine not only the kinematic growth of magnetic energy, but also the nonlinear mode couplings responsible for the migration of magnetic power across scales and the eventual saturation of the dynamo.

We show, both analytically and numerically, that in the kinematic limit with $\thkpq=\tau$, the EDQNM equations recover the classical kinematic dynamo descriptions of \citep{kazantsev, KN1967, KA92}. We then show that nonlinear magnetic-mode couplings provide a hyperdiffusive sink of magnetic energy (ME) at large scales and a net dissipative sink at small scales, with the former being mathematically similar to the result of \citep{Subramanian2003Hyperdiffusion}. Alongside velocity--magnetic interactions that amplify ME, nonlinear feedback produces Alfv\'enisation at small scales. Together, these processes shift the ME spectral peak or integral scale from resistive scales toward larger scales and lead to scale-by-scale saturation.

These theoretical insights motivate a broader exploration of the SSD across a wide parameter space using the full nonhelical EDQNM equations, utilizing the scale-dependent eddy damping time defined in Eq.~\eqref{eq:cl5} rather than the limiting case of $\thkpq = \tau$. By employing a logarithmic discretization grid, we achieve immense computational savings compared to DNS, allowing for a comprehensive study across extreme values of $\Rm$ and $\Rey$. 
Our primary findings are as follows:
\begin{itemize}
    \item \textbf{Critical magnetic Reynolds number:} 
    $\Rmcr$ is minimized near $\Pm=1$, increases toward an asymptotic regime for $\Pm<1${, turbulent flows}, and 
    {keeps on increasing} for $\Pm>1${, viscous flows}. 
    \item \textbf{Kinematic growth rate:} 
    The kinematic growth rate is controlled mainly by $\Rey$ for $\Pm>1$, where stretching is dominated by viscous-scale eddies, and mainly by $\Rm$ for $\Pm<1$, where resistive-scale eddies control the stretching. {When the flow is highly turbulent with parameters larger than $\Rm = \Rmasym \sim 10^6$ for $\Pm \leq 1$ and $\Rey = \Reasym \sim 10^6$ for $\Pm > 1$},
    $\gamma$ approaches an approximately $\Pm$-independent asymptotic regime.

    \item \textbf{Saturation efficiency:} 
    The saturation efficiency $\chi=\bsat^2/\usat^2$ increases with $\Rm$ for $\Pm<1$, varies more weakly for $\Pm>1$, and approaches an approximately $\Pm$-independent asymptotic regime at large $\Rey$.

    \item \textbf{Integral scale ratio:} 
    The ratio $k_M/k_V$ increases with $\Rm$ during the kinematic stage, but becomes remarkably constrained at saturation to a value of $3$, suggesting that nonlinear evolution limits the separation between the magnetic and kinetic integral scales.

    \item \textbf{Spectral behavior:} 
    For $\Pm>1$, magnetic energy dominates the kinetic energy at small scales and can transfer energy back to the velocity field, leading to small-scale Alfv\'enisation and aiding scale-by-scale saturation. In sufficiently turbulent regimes, the saturated KE and ME spectra approach each other more closely, suggesting a reduced sensitivity to the microscopic value of $\Pm$, reducing the system to a $\Pm = 1$-type state. {But the dissipative cutoff wavenumbers of the spectra increases with $\Rm$.}
\end{itemize}

These findings have direct astrophysical implications. Systems such as elliptical galaxies and the intracluster medium lack significant differential rotation. Thus, the small-scale dynamo is a leading candidate to explain their observed magnetization. The asymptotic states and ratios discovered here are useful for interpreting the magnetic energy and coherence scales observed in these environments \citep{PB_Faradayrot,Seta_2021_ellgal, GhoshBhat2025, Tevlin2025, Berlok_PICO2026}.
Additionally, these results have implications for interactions between the nonlinear large-scale dynamo and small-scale dynamo in turbulent systems \citep{BSB2016}.  

While EDQNM is a computationally efficient technique for modeling spectral evolution, it has limitations. The closure is {primarily} built upon the assumption of strict {spectral} isotropy. 
While justifiable in the kinematic stage, this may not strictly hold during the nonlinear saturation of the SSD. However, we argue that this assumption remains reasonable, because field stretching is dominated by the stronger, large-scale eddies, relative to which the small-scale magnetic field appears nearly isotropic. Furthermore, EDQNM cannot capture spatial intermittency, which requires high-resolution DNS. Finally, while EDQNM predicts enhanced nonlinear dissipation, it is difficult to map this with the magnetic reconnection events. 
Consequently, while the existence of these asymptotic states is robust, their exact numerical values may differ slightly in future high-resolution DNS studies. Nonetheless, these results represent a critical first-order calculation that paves the way for fully resolved simulations. 
{Moreover, the emergence of these asymptotic regimes may point to an underlying universality in the nonlinear dynamo saturation process, although establishing this will require a broader exploration of parameter space.}

\section{Data Availability}
The code used to solve the EDQNM equations, the resulting data, analysis scripts, and associated plots and videos are publicly available at \cite{data_edqnm}.

\section*{Acknowledgments}\label{sec:acknowledgments}
We thank Annick Pouquet, Sugan Durai Murugan, Vishal Vasan and the
members of ICTS plasma astrophysics group,  for useful discussions. 
This research was supported by the Department of Atomic Energy, Government of India, under Project
No. RTI4019. M.I.P. acknowledges the warm hospitality of
Inter-University Centre for Astronomy and Astrophysics
(IUCAA).

\bibliographystyle{apsrev4-2} 
\bibliography{reference}

\appendix

\begin{center}
\textbf{\large Appendix}
\end{center}

\section{Evaluation of transfer integrals}
\label{eval-tfr}

\subsection{Stretching term}\label{sec:tmvm}
Here we give the detailed evaluation of the transfer integrals in the case of $\thkpq = \tau = \const$. The integral over $\TMVM$ is
\begin{align}\label{eq:a0}
     I = \tau \int_{\Delta_k} \dd p\ \dd q\ \Bigg\{\frac{k^5}{p^3q}\ c_{kpq}\ E_p M_q &+ \frac{k^3}{pq}\ \hkpq\ M_p E_q \notag\\
    & - \frac{kp}{q}\ \hkpq\ E_q M_k\Bigg\}, \\
    I = I_1 + I_2 + I_3. \notag
\end{align}
The 3rd term is
\begin{align}\label{eq:a1}
    I_3 = -\tau k M_k \int_{\Delta_k} \dd p\ \dd q\ \frac{p}{q}\ \hkpq\ E_q.
\end{align}
From Eqs~\eqref{eq:cl3} and \eqref{eq:cl4} we have $\hkpq = \sin^2\beta$. And we transform the integration variable from $p$ to $\beta$ using
\begin{equation}\label{eq:a2}
    \dd p = \dd\beta\ \left(\frac{\p p}{\p\beta}\right) = \frac{kq}{p}\ \sin\beta\ \dd\beta,
\end{equation}
where we used the cosine identity
\begin{equation}\label{eq:a3}
    p^2 = k^2 + q^2 - 2kq \cos\beta. 
\end{equation}
Now we get the turbulent diffusion mentioned in Eq.~\eqref{eq:cl7b} of the main text
\begin{align}\label{eq:a4}
   I_3 &= -\tau k^2 M_k \int_0^\pi \dd\beta\;\sin^3{\beta} \int E_q \dd q, \\
       &= - \tfrac{4}{3} \tau k^2 M_k \int E_q \dd q.
\end{align}
Substitute Eqs~\eqref{eq:cl3}, \eqref{eq:cl4} and \eqref{eq:a2} in $I_1$ to get
\begin{align}\label{eq:5}
    I_1 &=  \tau \int_{\Delta_k} \dd p \; \dd q \;\frac{k^5}{p^3q} c_{kpq} \;E_p M_q, \notag\\
    &= \tau k^5 \int_0^\pi \sin^3\beta\ \dd\beta\ \int \dd q \;\cos\gamma\; \frac{E_p}{p^3} M_q, \notag\\
    &= \tau k^4 \int_0^\pi\sin^3\beta\ \dd\beta \int \dd q\ M_q E_p\ \frac{\left(k^2 + p^2 - q^2\right)}{2p^4},
\end{align}
in the final step we used the cosine identity,
\begin{equation}\label{eq:a6}
    q^2 = k^2 + p^2 - 2kp \cos\gamma. 
\end{equation}
To simplify the calculation we use $\beta \to \theta$ and $q\to x$, 
\begin{align}
    I_1 &= \tau k^4 \int_0^\pi\sin^3\theta \;\dd\theta\int \dd x\ M_x E_p\ \frac{\left(k^2 + p^2 - x^2\right)}{2p^4},\label{eq:a6a}
\end{align}
where
\begin{align}
    p &=  \left(k^2 + x^2 - 2kx\cos\theta\right)^{1/2}. \label{eq:a6b}   
\end{align}
The second integral is
\begin{align}\label{eq:a7}
    I_2 &= \tau k^3 \int_{\Delta_k} \frac{\dd p \; \dd q}{pq}\ h_{kpq}\ M_p E_q.
\end{align}
Note that the integration in Eq.~\eqref{eq:kin2} is over $M_k$, so we want to keep the measure $\dd p$. Transforming  
\begin{equation}\label{eq:a7a}
    \dd q = \frac{kp}{q}\ \sin\gamma\ \dd\gamma,
\end{equation}
from Eq.~\eqref{eq:a6} and combine with the sine formula,
\begin{equation}\label{eq:a8}
    \sin\beta = \frac{p}{q}\ \sin\gamma,
\end{equation}
to rewrite Eq.~\eqref{eq:a7} as
\begin{align}
    I_2 &= \tau k^4 \int_0^\pi\dd \gamma\ \sin^3\gamma\ \int \dd p\ M_p\ E_q\ \frac{p^2}{q^4}.
\end{align}
Use $\gamma \to \theta$ and $p\to x$ such that $q = \left(k^2 + x^2 - 2kx \ \cos\theta\right)^{1/2} = p$. Then we have
\begin{align}\label{eq:a9}
    I_2 &= \tau k^4 \int_0^\pi\dd \theta\ \sin^3\theta\ \int \dd x\ M_x\ E_q\ \frac{x^2}{q^4}.
\end{align}
Add eqs.~\eqref{eq:a6a}, \eqref{eq:a9} and use Eq.~\eqref{eq:a6b} to get
\begin{align}\label{eq:a10}
    I_1 + I_2 &= \int K_m(k,x) M_x\ \dd x,
\end{align}
where
\begin{align}\label{eq:a10a}
    K_m (k,x) = \tau k^4\int \dd\theta \ \sin^3\theta\ \left(\frac{k^2 + x^2 - kx\cos\theta}{q^4}\right) \ E_q,
\end{align}
as claimed in Eqs~\eqref{eq:cl7d} and \eqref{eq:cl7} with $x$ as the dummy variable.

\subsubsection{Equivalence with Kazantsev (1967)}\label{subsec:kaz}

In the $k-$space, \cite{kazantsev} derives the equation for 3D magnetic spectra \(G(k,t)\) as (see Eq.~(6.22) of \cite{shukurovkandubook} as well)
\begin{align}\label{eq:kaz_3d}
    \left(\frac{\partial}{\partial t} + 2\beta k^2\right) G(k,t) &= \int d^3\bm{q} \, G(p,t) V(q)\ \times \nonumber
    \\
    & \quad \quad \left(k^2 - \frac{(\bm{k}\cdot\bm{q}) (\bm{k}\cdot\bm{p}) (\bm{p}\cdot\bm{q})}{p^2 q^2}\right),
\end{align}
where $\bm{k} = \bm{p} + \bm{q}$. The 3D magnetic spectrum $G(k,t)$ and the kinetic spectrum $V(k)$ are defined via their respective ensemble averages
\begin{align}
    \langle b_i(\bm{k}, t) b_j(\bm{k}',t) \rangle &= G(k,t) \left(\delta_{ij} - \frac{k_i k_j}{k^2}\right) \delta(\bm{k} + \bm{k}'), \label{eq:def_G} \\
    \langle v_i(\bm{k}, t) v_j(\bm{k}',t') \rangle &= V(k)\, \delta(t-t') \left(\delta_{ij} - \frac{k_i k_j}{k^2}\right) \delta(\bm{k} + \bm{k}'). \label{eq:def_v}
\end{align}
Finally, the effective diffusivity is given by $\beta = \eta + T_L(0)$, where $T_L(0)$ is defined by \cite{shukurovkandubook}
\begin{align}\label{eq:def_TL}
    T_L(0) &= \frac{1}{3} \int_0^\infty dt' \langle \bm{v}(\bm{x},t) \cdot \bm{v}(\bm{x},t') \rangle \notag \\
    &= \frac{1}{3} \int_0^\infty dt' \langle \bm{v}(t)^2\rangle f(t - t') = \frac{1}{3} \tau \langle {\bm{v}}^2 \rangle.
\end{align}
Note that we drop the dependence on $\bm{x}$ based on the assumption of homogeneous turbulence and we defined the correlation time of the flow using $\tau = \int_0^\infty \dd t'\ f(t - t')$, where $f$ is an appropriate function determined from the flow.
The 1D magnetic energy spectrum $M(k,t)$ is obtained by 
\begin{align}\label{eq:def_M}
    M(k,t) = 4\pi k^2 G(k,t).
\end{align}
There is ambiguity in writing a similar expression relating the 1D kinetic energy spectra $E(k)$ from its 3D spectrum $V(k)$ due to the involvement of $\updelta(t-t')$ giving an extra time dimension. 
To fix this, we use the fact that inverse Fourier transform of $2V(k)$ is the contracted turbulent diffusion tensor in real space $T_{ii}(r)$, where 
\begin{align}
 \langle v_i (\bm{x}, t) v_j (\bm{x} +\bm{r}, t')\rangle &= T_{ij}(r) \delta(t-t'),\quad r = |\bm{r}|.
\end{align}
Then we obtain 
\begin{align}
    \langle v_i (\bm{r}, t)v_i (\bm{r}', t')\rangle = \int \dd^3 \bm{k}\ 2V(k) \delta (t - t') e^{\iota \bm{k}\cdot(\bm{r} - \bm{r}')}.
\end{align}
Consider the limit of $\bm{r} \to \bm{r}'$ and integrate over $t'$ from $-\infty$ to $\infty$ to obtain
\begin{align}
    2\tau\ \langle v(\bm{r}, t)^2\rangle = \int \dd^3 \bm{k}\ 2V(k).
\end{align}
Note that here we got $2\tau$, because the correlation time of the flow $\tau$ is defined through the integral over $t'$ from $0$ to $\infty.$ Comparing with the definition of 1D spectra
\[
\langle v^2 \rangle = 2\int E_k \dd k,
\]
yields the relation
\begin{align}\label{eq:KE31}
    E_k = \frac{2\pi k^2 V_k}{\tau},
\end{align}
between the 1D and 3D kinetic energy spectra.

We now address the integral on the RHS of Eq.~\eqref{eq:kaz_3d}:
\begin{align}\label{eq:rhs_kaz3d}
    I = \int d^3\bm{q} \, G(p,t) v(q) \left[ k^2 - \frac{(\bm{k}\cdot\bm{q})(\bm{k}\cdot\bm{p})(\bm{p}\cdot\bm{q})}{p^2 q^2} \right],
\end{align}
with $\bm{k} = \bm{p} + \bm{q}$. We transform the integration variable to $\bm{p}$, such that $d^3\bm{q} = d^3\bm{p} = p^2 \sin\theta \, dp \, d\theta \, d\phi$, where we take $\theta$  to be the angle between $\bm{k}$ and $\bm{p}$. From the triangle geometry, we have the relations $q^2 = k^2 + p^2 - 2kp\cos\theta$, $\bm{k}\cdot\bm{p} = kp\cos\theta$, $\bm{k}\cdot\bm{q} = k^2 - kp\cos\theta$, and $\bm{p}\cdot\bm{q} = kp\cos\theta - p^2$. Using these we simplify the expression in the bracket of Eq.~\eqref{eq:rhs_kaz3d} to
\begin{align}
    \bigg[\cdots\bigg] = \frac{k^2}{q^2} \sin^2\theta \left( k^2 + p^2 - kp\cos\theta \right).
\end{align}
Substituting this give
\begin{align}
    I &= \int_0^{2\pi} d\phi \int_0^\infty dp \int_0^\pi d\theta \, p^2 \sin\theta \left( \frac{M_p}{4\pi p^2} \right) \left( \frac{\tau E_q}{2\pi q^2} \right) \nonumber \\
    & \quad \quad \quad \quad \quad \qquad \quad \left[ \frac{k^2}{q^2} \sin^2\theta \left( k^2 + p^2 - kp\cos\theta \right) \right] \nonumber \\
    &= \frac{\tau k^2}{4\pi} \int_0^\infty dp \, M_p \int_0^\pi d\theta \, \sin^3\theta \left( k^2 + p^2 - kp\cos\theta \right) \frac{E_q}{q^4}.
\end{align}
To rewrite the LHS in terms of 1D ME spectra we divide it by $4\pi k^2$ and after rearranging we obtain
\begin{equation}
    \begin{split}
         \left(\frac{\partial}{\partial t} + 2\beta k^2\right) M(k,t) = \int K_m (k,x) M_x \, \dd x, \\
    K_m(k,x) = \tau k^4 \int_0^\pi d\theta \, \sin^3\theta \left( k^2 + p^2 - kp\cos\theta \right) \frac{E_q}{q^4},
    \end{split}
\end{equation}
as claimed in the Eqs~\eqref{eq:kin2}~-~\eqref{eq:kin3}.

The time evolution of $M(k,t)$ according to \cite{kazantsev} is (see Eq.~(33))
\begin{align}\label{eq:kzgr1}
    M(k,t) \propto \exp{\left(\frac{1}{5} |\epsilon| v_2 t\right)},
\end{align}
where $\epsilon$ lies in the narrow range (see Eq.~(30))
\[
-4 < \epsilon < \epsilon_0 +\updelta,\quad \updelta \ll 1,
\]
with (see Eq.~(9) of \cite{kazantsev})
\[
v_2 = \frac{1}{6} \int \dd^3 \bm{k}\ k^2 V(k) =  \frac{\tau}{3} \int \dd k\ k^2 E_k,
\]
where we used Eq.~\eqref{eq:KE31}.
The minimum growth rate of magnetic energy is for $|\epsilon| = 15/4$, thus $E_M \propto \exp{\left(3 \gamma t /4\right)}$, with
\begin{equation}\label{eq:ka6}
    \gamma = \frac{\tau}{3} \int \dd k\ k^2 E_k.
\end{equation}
As shown in the main text, this gives a decent fit in simulations of kinematic kazantsev dynamo with $\tau = 1$ (see Fig.~\ref{fig:Emag_lin512}).

\subsubsection{Equivalence with Kulsrud \& Anderson (1992)}
The evolution of magnetic energy spectra is given by the Eqs~(2.30 - 2.31) of \cite{KA92}, 
\begin{align}
    \frac{\p M}{\p t} &= \int K_m(k,x)M(x)\dd x - 2k^2 \frac{\eta_T}{4\pi}M(k) - 2k^2 \frac{\eta_s}{4\pi}M(k),\label{eq:ka1a}\\
    K_m (k,x) &= 2\pi k^4 \int \dd\theta \ \sin^3\theta\ \left(\frac{k^2 + x^2 - kx\cos\theta}{q^2}\right) \ U(q),\label{eq:ka1b}   
\end{align}
with $q^2 \equiv k^2 + x^2 - 2kx\cos{\theta}$. Note that the ${1}/{4\pi}$ with $\eta_s$ (Spitzer resistivity) and $\eta_T$ are because in their definition, magnetic diffusivity which is the coefficient of $\nabla^2\bm{B}$ in the induction equation appears with the same factor. So we can identify $\eta \equiv {\eta_s}/{4\pi}$ and $\eta_T \equiv {\eta_T^{KA}}/{4\pi}$. 
And $U(q)$ is defined in Eq.~(2.38) through the correlation tensor of Fourier turbulent velocities,
\begin{align}\label{eq:ka2}
    \langle\bm{v}^*_{\bm{k}'}(t')\ \bm{v}_{\bm{k}}(t)\rangle
     = \left(\frac{2\pi}{L}\right)^3 U(k) (\bm{I} - \hat{\bm{k}}\hat{\bm{k}}) \ \updelta_{\bm{k}',\bm{k}}\ \updelta(t'-t),
\end{align}
where $\bm{I}$ is the unit dyadic and $\hat{\bm{k}}$ is the unit vector along $\bm{k}$ and we consider only the nonhelical motions. The factors such as $\left({L}/{2\pi}\right)^3$ is due to the Fourier transform conversion in the discrete space. Similar to the Kazantsev formulation, we use Eq.~\eqref{eq:KE31} to transform from 3D to 1D kinetic spectra.
The turbulent diffusivity is given in Eq.~(2.36)
\begin{align}\label{eq:ka4}
    \frac{\eta_T}{4\pi} = \frac{1}{3}\int U(q) \dd^3 q = \frac{2\tau}{3}\int E_k\, \dd k = \frac{1}{3}\langle \bm{v}^2\rangle.
\end{align}
Similarly rewriting Eq.~\eqref{eq:ka1b} in terms of $E_k^{KA}$ we get
\begin{align}\label{eq:ka5}
    K_m (k,x) &= \tau k^4 \int \dd\theta \ \sin^3\theta\ \left(\frac{k^2 + x^2 - kx\cos\theta}{q^4}\right) \ E_q. 
\end{align}
Thus this formulation is equivalent to that in \cite{kazantsev} and presented here. 

\subsubsection{Back reaction on magnetic energy spectrum}\label{sec:tmmm}
Using Eqs~\eqref{eq:cl3} and \eqref{eq:cl4}, integral over $\TMMM$ is
\begin{align}\label{eq:a11}
    I &= -\tau k^2 M_k \int_{\Delta_k} \dd p\ \dd q\ z(1-y^2)\ \frac{M_q}{q}.
\end{align}
Use Eqs~\eqref{eq:a2}, \eqref{eq:a3} and triangle identity
\begin{equation}\label{eq:a11a}
    z = \cos\gamma = (k - q\cos\beta)/p
\end{equation}
to get
\begin{align}\label{eq:a12}
    I &= -\tau k^3 M_k\int\dd q\ M_q \int_0^\pi \dd \beta\ \sin^3\beta\ \frac{k-q\cos\beta}{\left(k^2 + q^2 - 2kq\cos\beta\right)},
\end{align}
To perform the integral over $\beta$, set $a = k^2 + q^2,\ b = 2kq, \ x = \cos\beta,$ such that
\begin{align}
    \int_{-1}^1 \dd x\ \frac{(k - qx)(1-x^2)}{a - bx} 
    = \frac{2ak}{b^2} + \frac{4q}{3b} - \frac{2a^2 q}{b^3} \notag \\[10pt]
    \quad + \left(1 - \frac{a^2}{b^2}\right) \left(\frac{k}{b} - \frac{aq}{b^2}\right) \ln{\left( \left| \frac{a+b}{a-b} \right| \right)} \notag\\
    = \frac{2}{3k} + \frac{k^4 - q^4}{4k^3 q^2} - \frac{\left(k^2 - q^2\right)^3}{8q^3 k^4} \ln{\left( \left| \frac{k+q}{k-q} \right| \right)}.
\end{align}
Substituting in Eq.~\eqref{eq:a12} gives
\begin{align}\label{eq:a13}
    I &= -\frac{2}{3}\tau\ k^2 M_k \left(\int M_q \dd q\right) - \frac{1}{4}\tau\ k^4 M_k \left(\int \dd q \ \frac{M_q}{q^2}\right) 
      \notag\\
      &+ \frac{1}{4}\tau\ M_k \left(\int \dd q\ q^2 M_q\right)
      \\
      &+ \tau M_k \int \dd q\ M_q \ \frac{\left(k^2 - q^2\right)^3}{8 k q^3} \ln{\left(\bigg|\frac{k+q}{k-q}\bigg|\right)}\notag
      \\
      &= I_1 + I_2 + I_3 + I_4.
\end{align}
The first three terms correspond to nonlinear diffusion, hyper-diffusion and a growth due to nonlinear stretching respectively. To study the final term $I_4$ we consider the small scale and large scale limits using the expansion
\[
\left(1-x^2\right)^3 \ln{\left(\frac{1+x}{1-x}\right)} \simeq 2 \left(x - \frac{8}{3}x^3 + \frac{11}{5}x^5\right) + \mathcal{O}(x^7).
\]
On large scales ($x = k/q \ll 1$), 
\begin{align}\label{eq:a14a}
  I_4 &\simeq \frac{2}{3}\tau k^2 M_k\ \left(\int \dd q\ M_q\right) - \frac{1}{4}\tau\ M_k \left(\int \dd q\ q^2 M_q\right) \notag\\[3pt]
  &- \frac{11}{20}\tau k^4 M_k \left(\int \dd q\  \frac{M_q}{q^2}\right).
\end{align}
In this limit Eq.~\eqref{eq:a13} simplifies to
\begin{align}\label{eq:a14}
    I &= - \frac{4}{5}\tau\ k^4 M_k \left(\int \dd q \ \frac{M_q}{q^2}\right).
\end{align}
Thus the back reaction from magnetic field leads to a hyper diffusion of magnetic energy at large scales. 

On small scales ($x = q/k \ll 1$):
\begin{align}\label{eq:a15}
   I_4 \simeq &-\frac{2}{3}\tau k^2 M_k\ \left(\int \dd q\ M_q\right) 
     + \frac{1}{4}\tau\ M_k k^4 \left(\int \dd q\  \frac{M_q}{q^2}\right)
     \notag\\
     &+ \frac{11}{20}\tau M_k \left(\int \dd q\  q^2 M_q\right),
\end{align}
and Eq.~\eqref{eq:a13} becomes
\begin{align}\label{eq:a16}
    I &= -\frac{4}{3}\tau\ k^2 M_k \left(\int M_q \dd q\right) 
      + \frac{4}{5}\tau\ M_k \left(\int \dd q\ q^2 M_q\right),
\end{align}
a combination of nonlinear diffusion and stretching of magnetic field.

\subsubsection{Back reaction on kinetic energy spectrum}\label{sec:tvvm}
The integral of $\TVVM$ is calculated using Eqs~\eqref{eq:cl2}~-~\eqref{eq:cl3}, then transforming the integral using Eqs~\eqref{eq:a2} and \eqref{eq:a11a} gives
\begin{align}\label{eq:a17}
    \tau\intDk \dd p\ \dd q\ \TVVM 
    &= -\tau\ kE_k\int\dd q\ M_q \int_0^\pi \left(k-q\cos\beta\right)\sin^3\beta\ \dd\beta\notag\\
    &= -\frac{4}{3}\tau k^2 E_k \int\dd q\ M_q,
\end{align}
the nonlinear diffusion in the KE.

\subsection{Numerical setup}\label{num-setup}
\subsubsection{Discretization}
We employ two distinct wavenumber discretizations depending on the scenario. For the EDQNM simulations with scale dependent damping time $\thkpq$,  we utilize a logarithmically spaced grid:
\[
k_i = \kmin\ r^{i},\quad i = 0,1,\cdots,N-1,
\]
where $r = 2^{1/F_{\text{res}}}$ with the sub-octave resolution $F_{\text{res}}$. Throughout this work, unless otherwise stated, we use $F_{\text{res}} = 16$.

\begin{figure}
    \centering
    \includegraphics[width=\linewidth]{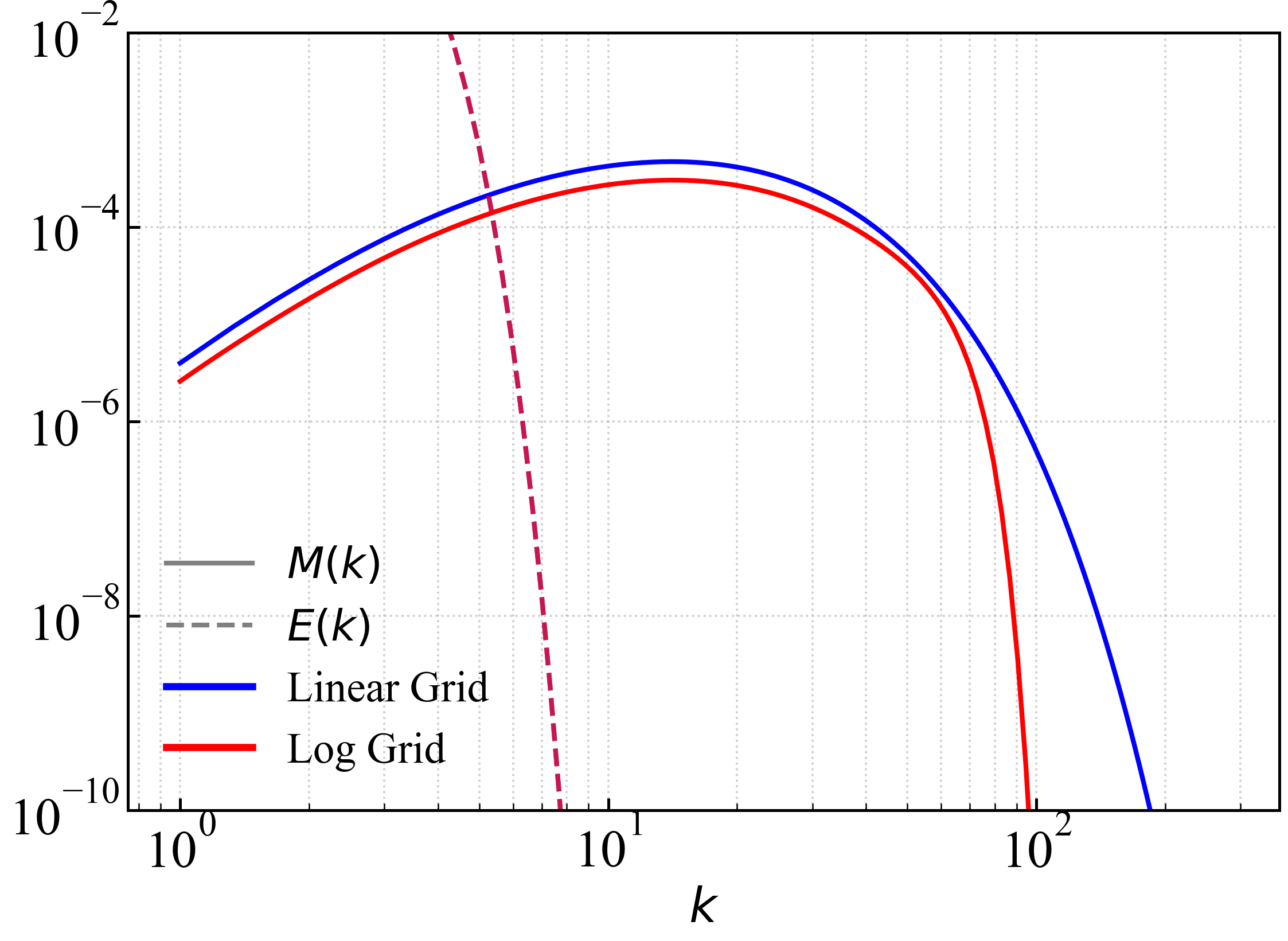}
    \caption{Magnetic and kinetic energy spectra in the case of kinematic Kazantsev dynamo given in the main text for wavenumber grids which are linear and logarithmically spaced. Notice the sharp fall in magnetic energy spectrum at small scales in case of logarithmic grid due to the lack of sufficient points there.}
    \label{fig:kinkaz_linlog}
\end{figure}

When studying the Kazantsev dynamo with $\thkpq = \tau = \const$, we utilize a linear grid:
\[
k_i = \kmin + i\ \Delta k,\quad i = 0,1,2,\cdots,N-1.
\]
This is because, in this case the logarithmic grid fails to capture the small-scale dynamics due to lack of sufficient points and leads to sharp fall in the ME spectrum (red solid line), as shown in Fig.~\ref{fig:kinkaz_linlog}. By using a linear grid we ensure sufficient modes at high wavenumbers and smooth fall off the spectra (blue solid line). But this increases the computational expense significantly and restricts us from exploring wide variety of parameters for the Kazantsev dynamo. 

\subsubsection*{Transfer terms}
Evaluating the double integral over the nonlinear transfer terms in Eqs~\eqref{eq:cl1} - \eqref{eq:cl1a} is the computationally dominant bottleneck of the simulation. For a given wavenumber $k$, the integration is strictly bounded by the geometric triangle condition $|k-p| \le q \le k+p$ \cite{Murugan2026}
\[
\int_{\Delta_k} \dd p \; \dd q \equiv \sum_{p=\kmin}^{\kmax} \Delta p\; \sum_{q = |k-p|}^{k+p} \Delta q.
\]
To optimize this, we precompute a matrix containing the discrete index bounds $[m_{\text{min}}(i,j), m_{\text{max}}(i,j)]$, corresponding to wavenumber $q$, for all valid triads, where $i, j$ denotes the wavenumbers $k$ and $p$ respectively. 
Since the native Python loop execution is prohibitively slow for large $N$, the RHS evaluations including the time-dependent relaxation time $\thkpq$ are explicitly parallelized and Just-In-Time (JIT) compiled using the \verb|Numba| library \cite{numba}. Time integration is performed using the implicit Backward Differentiation Formula (\verb|BDF|) scheme provided by \verb|scipy.integrate.solve_ivp| \cite{scipy}, with strict relative and absolute error tolerances set to \verb|rtol|$=10^{-6}$ and \verb|atol|$=10^{-10}$, respectively.

\begin{figure}
	\centering
	\includegraphics[width=\linewidth]{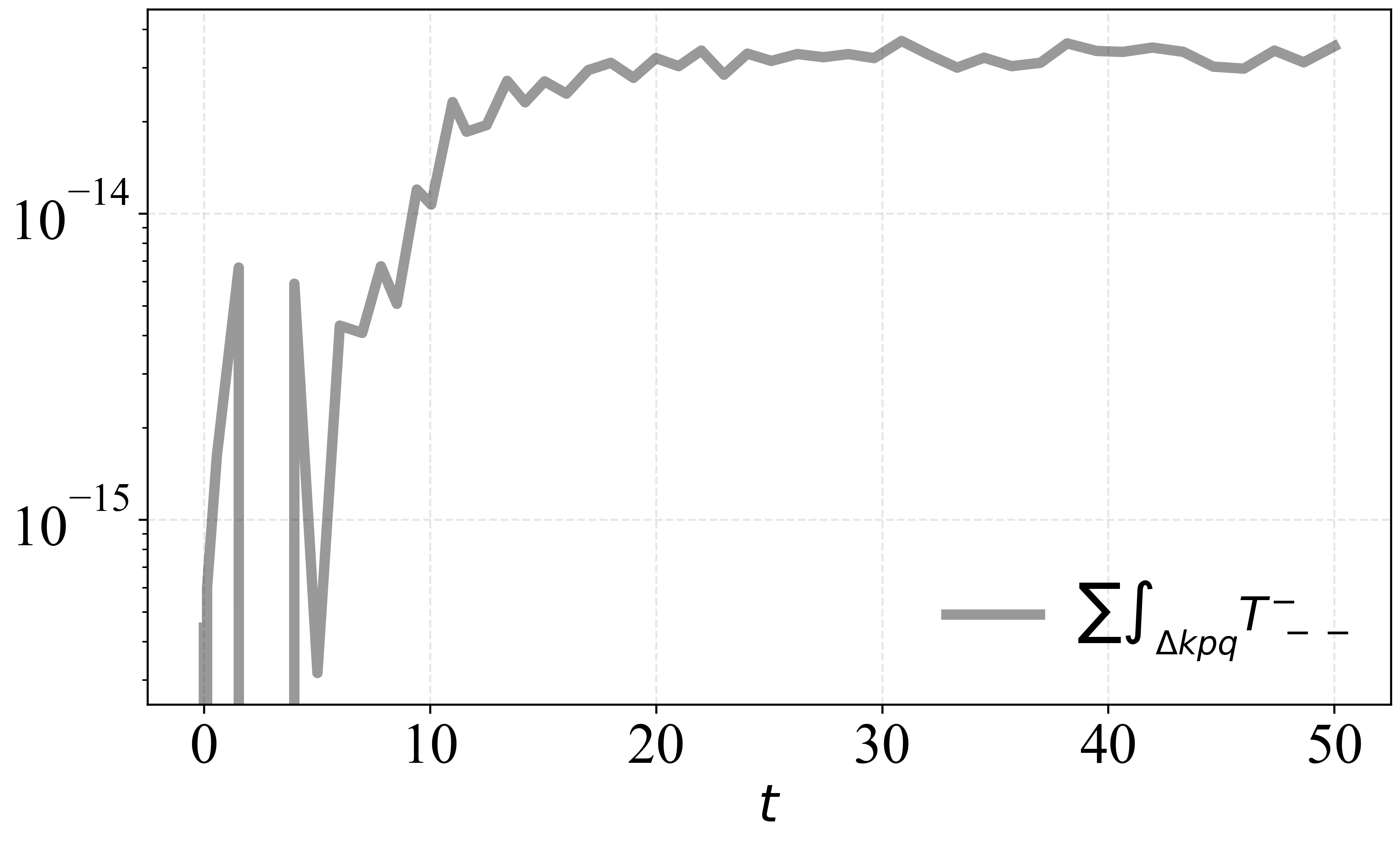}
	\caption{The sum of integrated transfer terms remains at machine zero precision throughout the kinematic simulation with $\thkpq = 1$, confirming energy conservation.}
	\label{fig:kazfull_check}
\end{figure}

\begin{figure}
	\centering
	\includegraphics[width=\linewidth]{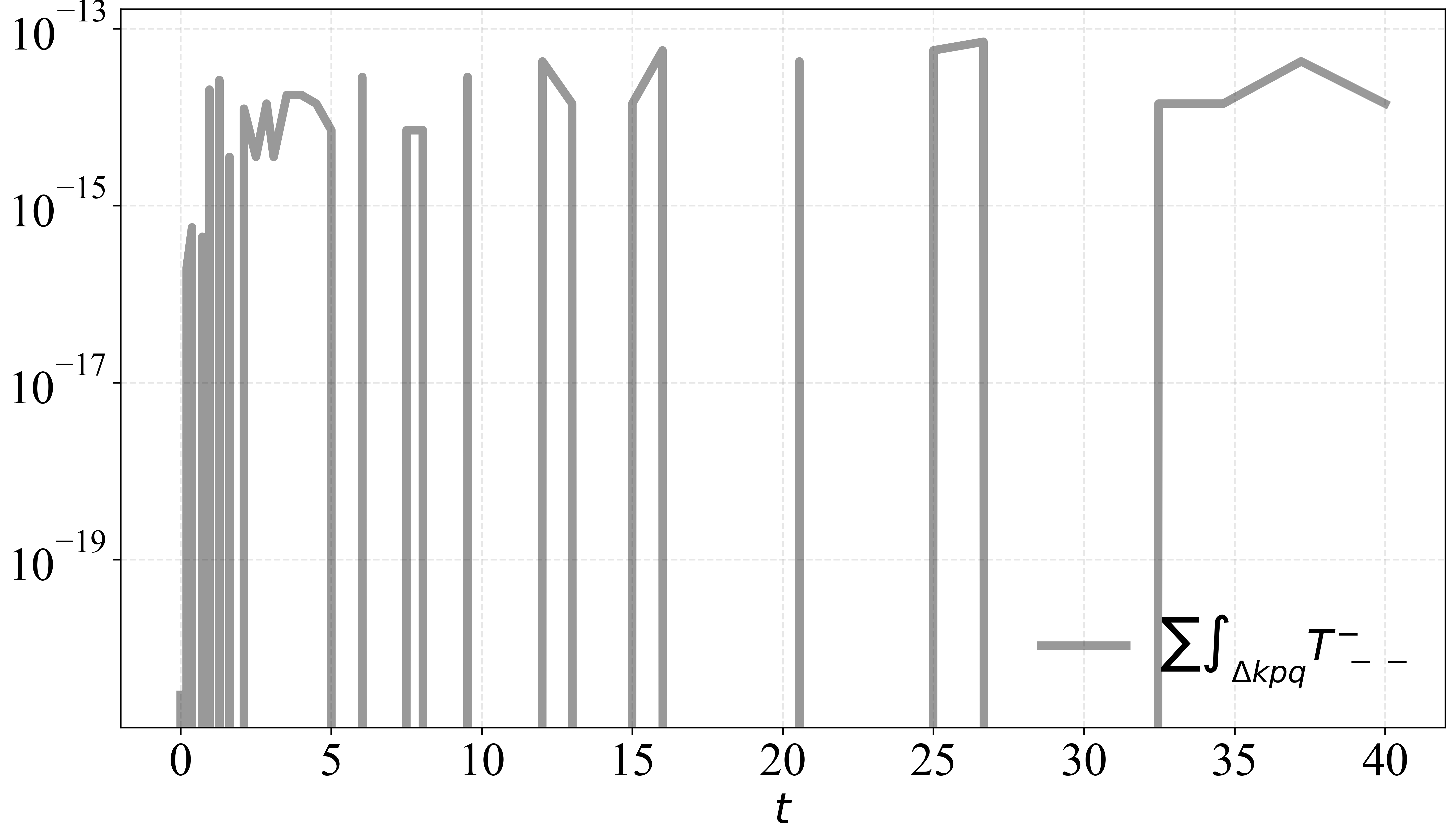}
	\caption{The sum of integrated transfer terms vanishes to machine precision throughout the full nonlinear EDQNM simulation.}
	\label{fig:edqnm_check}
\end{figure}

A rigorous check of our numerical implementation is the exact conservation of total energy.
This requires that the discrete sum over all transfer terms vanishes to machine precision at every time step
\[
\sum_{i=0}^{N-1} \Delta k_i \left( \int_{\Delta_k} \dd p\ \dd q \left[ \TVVV + \TVVM + \TVMM + \TMVM + \TMMM \right] \right) \simeq 0.
\]
Figures~\ref{fig:kazfull_check} and \ref{fig:edqnm_check} confirm the global energy conservation throughout the simulation for both the Kazantsev case ($\thkpq = 1$) and the full EDQNM.

\subsection{Convergence study}
\begin{figure}
    \centering
    \includegraphics[width=\linewidth]{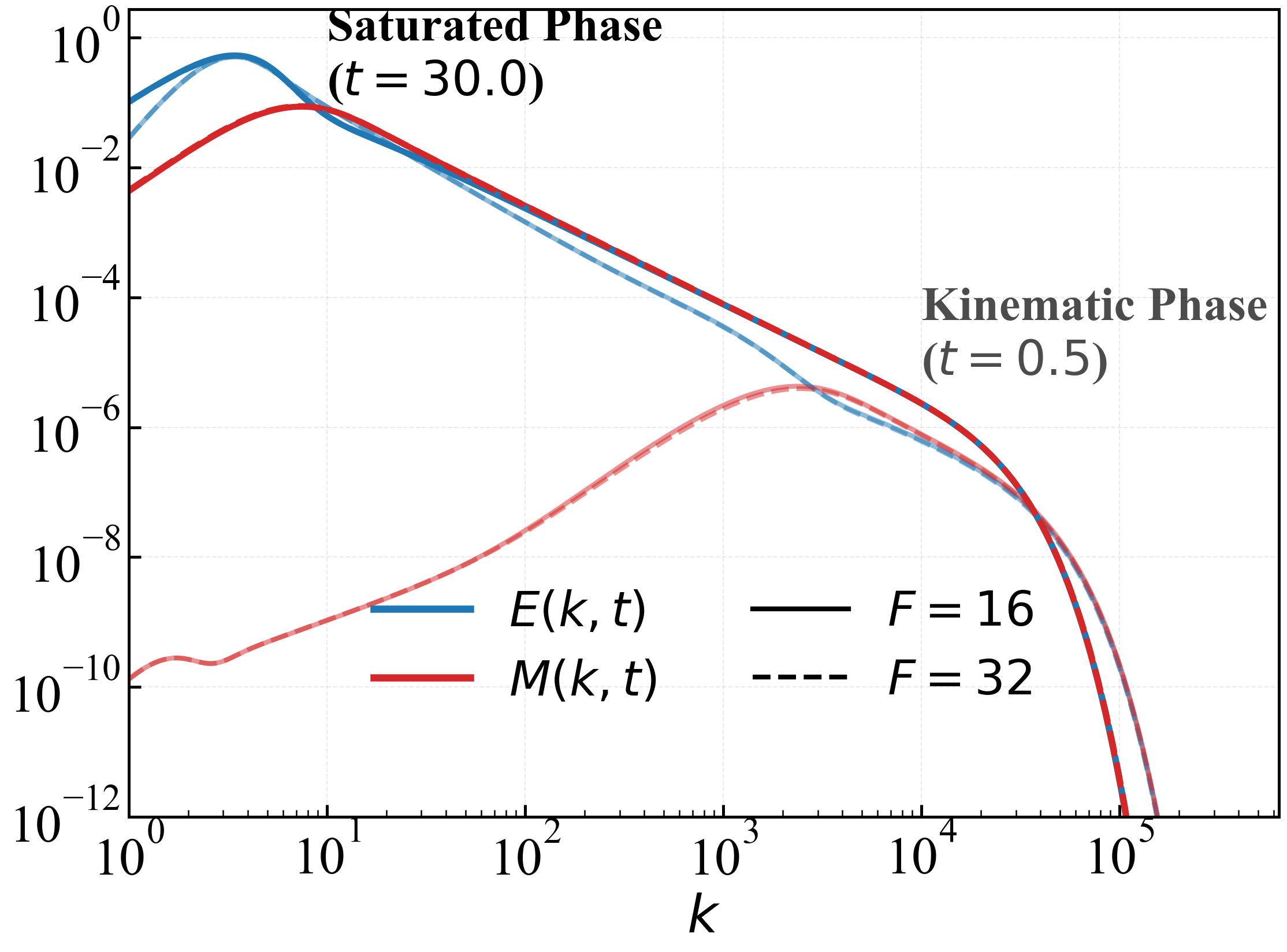}
    \caption{Magnetic and kinetic energy spectra in the kinematic (light solid line) and saturated (dark solid line) phases for a run with $\Pm=1,\, \Rm \approx 1.4 \times 10^6$. The excellent overlap between $F_{\text{res}}=16$ (solid lines) and $F_{\text{res}}=32$ (dashed lines) establishes numerical convergence.}
    \label{fig:convergence_edqnm_pm1}
\end{figure}

\begin{figure}
    \centering
    \includegraphics[width=\linewidth]{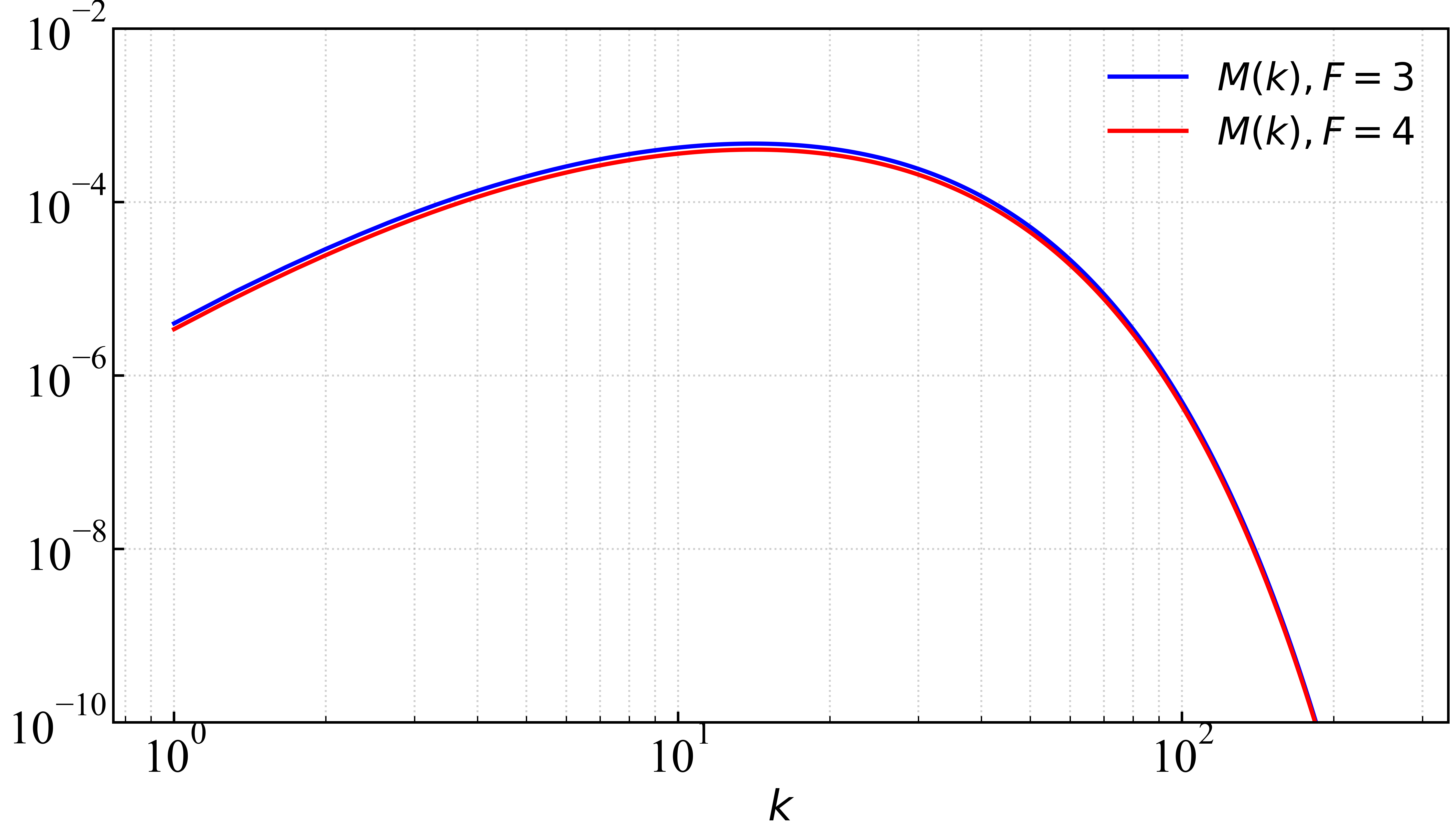}
    \caption{Magnetic energy spectra in the case of kinematic Kazantsev dynamo given in the main text for linear wavenumber grids with fineness of $F=3$ (blue) and 4 (red).}
    \label{fig:convergence_kinKaz}
\end{figure}

To verify numerical convergence and ensure the inertial and dissipative ranges are fully resolved, we repeated the target simulations with a doubled resolution of $F_{\text{res}} = 32$. As demonstrated in Fig.~\ref{fig:convergence_edqnm_pm1}, the kinetic and magnetic spectra during both the kinematic phase and the fully saturated nonlinear phase are indistinguishable between the two resolutions, confirming the convergence. Similarly we select the maximum wavenumber such that it is 4 times the $\text{max} (k_\eta, k_\nu)$ to ensure that the small scale dynamics is completely resolved. Similarly in the case of Kazantsev dynamo, we increase the fineness of the linear wavenumber grid from $F=3$ to 4 and find that the spectra lies decently close to one another as shown in Fig.~\ref{fig:convergence_kinKaz}.

\end{document}